\documentclass[reprint,amsmath,amssymb,aps]{revtex4-2}
\usepackage{graphicx}
\usepackage{amsmath}
\usepackage{bm}
\usepackage{hyperref}
\usepackage{xcolor}
\usepackage{float}
\usepackage{array}
\usepackage{helvet}
\begin{document}
\preprint{APS/123-QED}
	
	\title{Engineering nonlinear magnon scattering in artificial spin ice via vertex dipolar control}

\author{Waleed I. Waseer}
\affiliation{School of Physics and State Key Laboratory of Electronics Thin Films and Integrated Devices, University of Electronics and Science and Technology of China, Chengdu 610054, China}


\author{Peng Yan}%
\email{yan@uestc.edu.cn}
\affiliation{School of Physics and State Key Laboratory of Electronics Thin Films and Integrated Devices, University of Electronics and Science and Technology of China, Chengdu 610054, China}
	
	\begin{abstract}
Artificial spin ice (ASI), composed of geometrically frustrated arrays of interacting nanoislands, provides a versatile platform for reprogrammable magnonic functionality. However, the commonly used geometric control parameters such as island length, width, and aspect ratio simultaneously modify the island footprint, inter island dipolar spacing, and shape anisotropy, making it difficult to tune the nonlinear response independently of the linear spectrum and lattice density. Using micromagnetic simulations of kagome ASI under strong microwave drive, we identify edge curvature as a geometric degree of freedom that separates nonlinear magnon scattering from the island footprint. Sharp tipped islands predominantly generate integer harmonics, whereas dumbbell shaped tips produce a transition toward subharmonic rich spectra by concentrating demagnetizing and exchange fields near the island ends without changing the overall island volume or lattice spacing. By mapping the curvature and drive parameter space, we identify a continuous threshold for subharmonic onset controlled by tip curvature. We further show that the angular dependence of the second harmonic amplitude reverses between sharp and dumbbell geometries, providing an experimentally accessible signature of curvature localized nonlinearity. Width and leg length asymmetry can also modify harmonic amplitudes, but they do not remove the intrinsic trade off between footprint and coupling. These results establish tip curvature as a footprint preserving design parameter for engineering nonlinear magnon scattering in ASI, with implications for reconfigurable magnonic devices.

	\end{abstract}
	
	\maketitle
	
\section{\label{sec:level1} Introduction}
Artificial spin ice (ASI) is a two-dimensional lattice of interacting nanomagnets (``macrospins'') that mimics the frustration observed in water/ice \cite{harris1997geometrical,ramirez1994strongly,bramwell2001spin}. ASI's periodic nanomagnet arrangement offers a reconfigurable platform for magnonics and the underlying microstate governs spin-wave propagation, enabling ASI to function as a tunable magnonic crystal \cite{gliga2020dynamics,kaffash2021nanomagnonics,lendinez2019magnetization} for future re-programmable devices \cite{gliga2020dynamics,lendinez2019magnetization}. In recent years, researchers have achieved precise control over ASI microstates and explored their high-frequency dynamics, highlighting potential applications in dense memory and neuromorphic hardware \cite{budrikis2012network,heyderman2013artificial,nisoli2007ground,branford2012emerging}.

A key advantage of ASI is its ability to create an enormous variety of geometries, limited primarily by current nanofabrication techniques, which now include complex three-dimensional structures. Previous studies of square and kagome ASIs have examined bulk resonance modes tied to each sublattice's macrospin resonance, using techniques like Brillouin light scattering, micromagnetic simulations, and deliberate shape-anisotropy modifications to isolate different sublattice contributions \cite{sklenar2013broadband,carlotti2014micro,bang2019angular,shen2022dipolar}.

Researchers have also shown that topological defects formed during switching can reshape local magnon modes, and a distinct vertex localized mode arising from direct exchange between neighboring macrospins has been identified. Although it's known that dynamic behavior depends on the ASI microstate, the role of dipolar coupling in shaping these dynamics has been less explored. A recent bi-component square ASI study revealed that significant dynamic dipolar coupling can cause mode hybridization and anticrossings when sublattice modes become degenerate; rotating the applied field can tune this degeneracy \cite{gartside2021reconfigurable,dion2022observation}. However, reliably accessing a diverse set of microstates remains difficult. Current approaches (AC demagnetization, rotating-field sequences, or MFM-tip-induced reversal) are either stochastic or too slow for practical devices. To address this, Shen \emph{et al.} propose altering nanowire aspect ratios or shape anisotropy across sublattices trading away some degeneracy in order to gain deterministic microstate control through differing coercive fields \cite{shen2022dipolar}.

Despite substantial theoretical \cite{gliga2013spectral,jungfleisch2016dynamic,iacocca2016reconfigurable,iacocca2017symmetry} and experimental \cite{jungfleisch2016dynamic,zhou2016large,kempinger2021field,bhat2020magnon,bang2020influence,mamica2018spin,dion2019tunable,arroo2019sculpting,lendinez2021emergent,vanstone2022spectral,gartside2022reconfigurable} advances in probing ASI dynamics, most detected modes have been limited to the system's linear eigenfrequencies, typically accessed via microwave absorption measurements (using stripline antennas) or Brillouin light scattering of thermally populated magnons \cite{li2017thickness,lendinez2021observation}. Nonlinear processes such as three- and four-magnon scattering, which carry information about magnon coherence and lie at the heart of frequency mixing and parametric amplification, have been studied extensively in continuous ferromagnetic films and individual nanostructures \cite{wang2021magnonic,ulrichs2011linear,melkov2013nonlinear,guo2014parametric,haghshenasfard2017suhl}. Their counterpart in ASI was opened up by Lendinez \emph{et al.} \cite{lendinez2023nonlinear}, who experimentally demonstrated nonlinear multi-magnon scattering in ASI under strong microwave drive and showed that the lattice supports rich harmonic and subharmonic generation. This work establishes ASI as a genuinely nonlinear magnonic medium and raises an immediate follow up question that remains open: how can the strength and spectral content of these nonlinear processes be controlled by geometry?
The geometric knobs explored to date in ASI, namely island length, width, and aspect ratio \cite{shen2022dipolar,dion2019tunable,arroo2019sculpting}, all share an intrinsic limitation: they simultaneously vary the island footprint, the inter island dipolar spacing, and the shape anisotropy. Increasing island volume to enhance dipolar coupling, for instance, also reduces inter element separation and shifts the linear eigenfrequencies, so the nonlinear response cannot be tuned independently of the linear spectrum or the lattice density. A geometric degree of freedom that modifies nonlinear scattering without altering footprint or lattice spacing has not been identified, to the best of our knowledge.

In this work, we propose and demonstrate that tip curvature provides exactly such a knob. Using micromagnetic simulations of kagome ASI, we systematically vary nano island width, leg symmetry, and edge curvature to study their impact on ferromagnetic resonance and higher order harmonic generation. We first show that wider nano islands, with increased dipolar interaction volume, support stronger nonlinear spin wave responses, while geometrical asymmetries introduced by elongating one leg of the vertex create imbalanced dipolar fields that suppress coherent dynamics and reduce harmonic content. Both of these parameters alter the footprint and dipolar coupling simultaneously, thereby confirming, rather than circumventing, the limitation noted above. Against this baseline, we then introduce edge curvature in the form of dumbbell shaped tips. Dumbbell tips concentrate demagnetizing and exchange fields at the island ends and act as localized sources of nonlinearity, producing a qualitative transition from harmonic only spectra in sharp tipped islands to subharmonic rich spectra in dumbbell tipped islands, all without increasing the island footprint or reducing inter element spacing. Finally, we show that the angle of the applied excitation field provides a distinctive experimental fingerprint of this mechanism: the angular dependence of the second harmonic amplitude inverts between sharp and dumbbell geometries, offering a direct, falsifiable signature accessible to broadband ferromagnetic resonance measurements. Together, these results reveal how geometric design principles in ASI enable precise, footprint preserving control over nonlinear magnon processes, with implications for reprogrammable magnonic devices.

   \begin{figure}[!htph]
\centering
		\includegraphics[width=.9\linewidth]{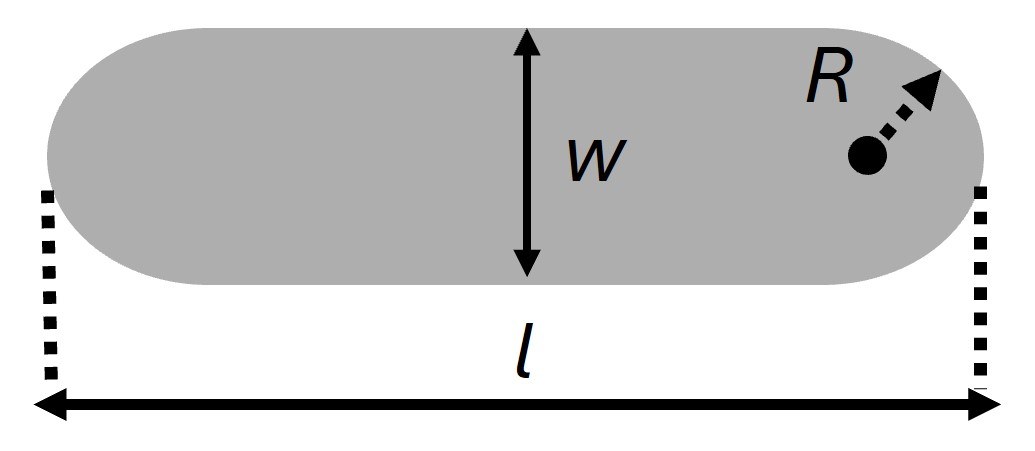}
	\caption{Schematic diagram of a single nano island.  Here $W$ indicate the width and $l$ is the total length of the nano island. $R$ is radius of semi circle.   }
	\label{fig1a}
\end{figure} 
\begin{figure*}[!htph]
    \centering

    \begin{tabular}{c c}
         \textsf{ (a) Hysteresis curve} & \textsf{(b) Resonance spectrum}  \\
        \includegraphics[width=0.45\linewidth]{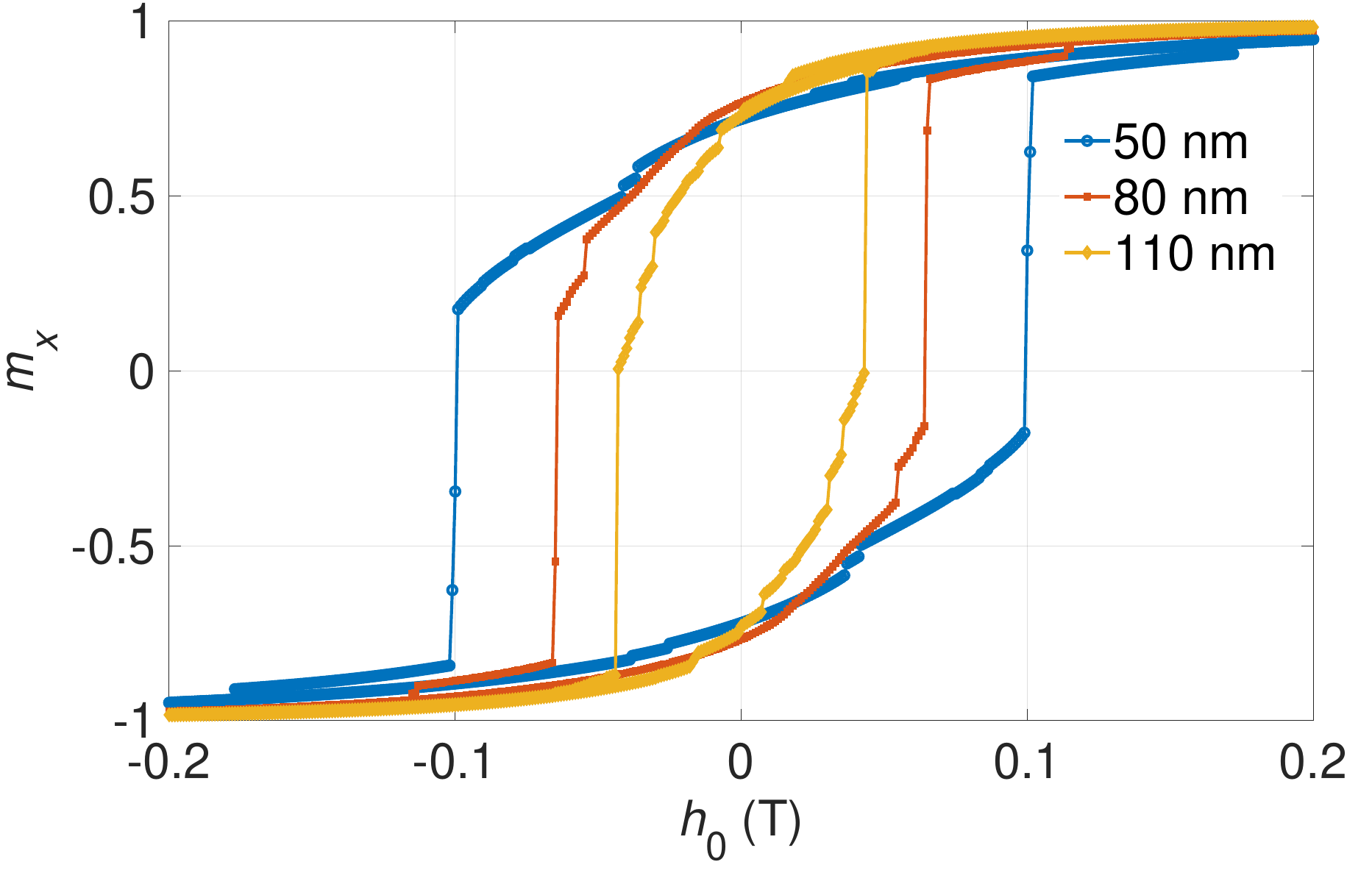} & 
        \includegraphics[width=0.45\linewidth]{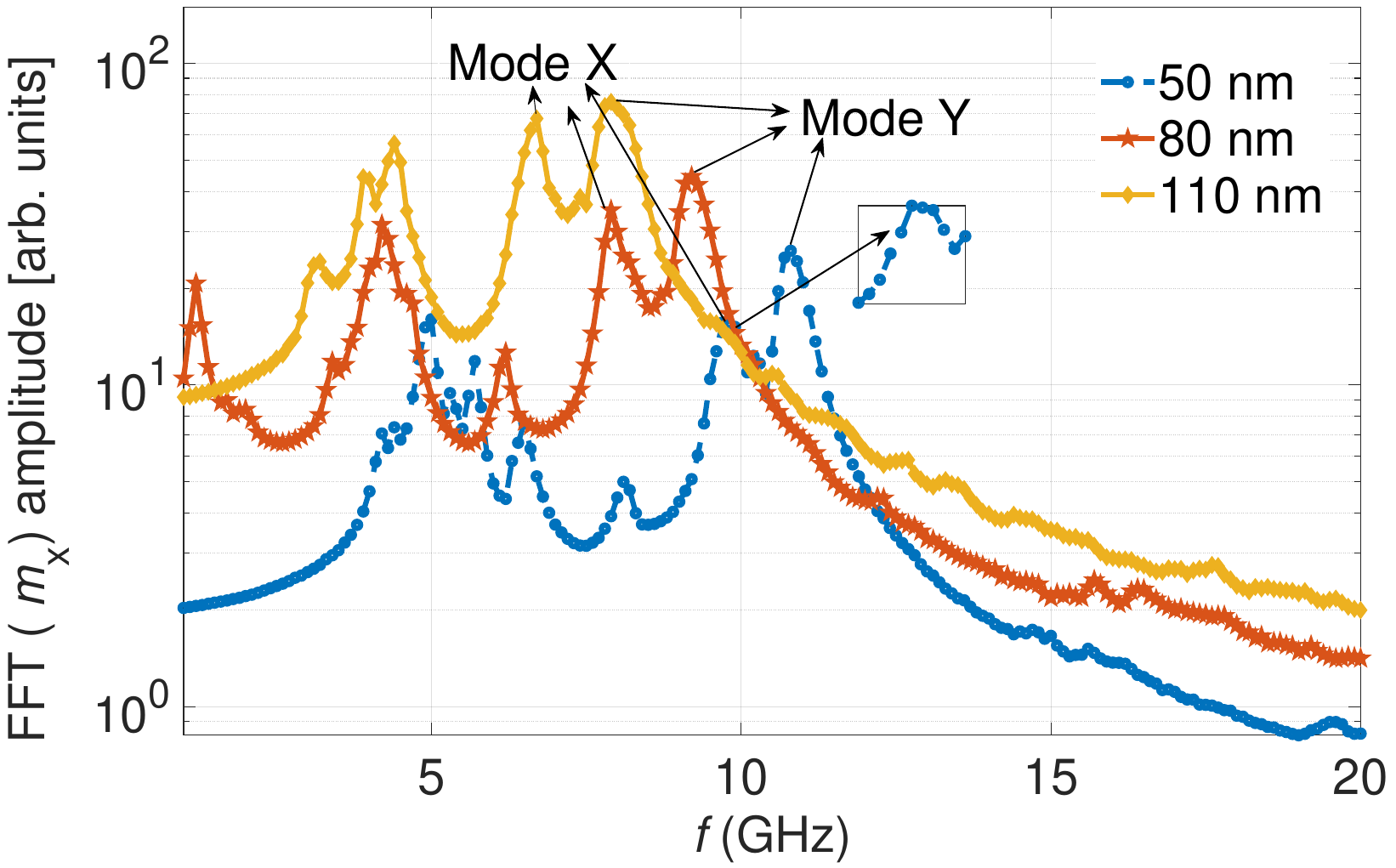} \\
  
    \end{tabular}

    \caption{Hysteresis loops (a) and resonance spectra (b) for nano islands of widths w = 50, 80, 110 nm and fixed length l = 260 nm.}
    \label{fig1}
\end{figure*}

\begin{table*}[!htph]
    \centering
    \vspace{1em} 
\centering
    \begin{tabular}{|c|c|c|c|}
        \hline
        Width & Geometry & Mode X & Mode Y \\
        \hline
     \centering  \raisebox{5\height}{50\,nm}    & \includegraphics[width=0.195\linewidth]{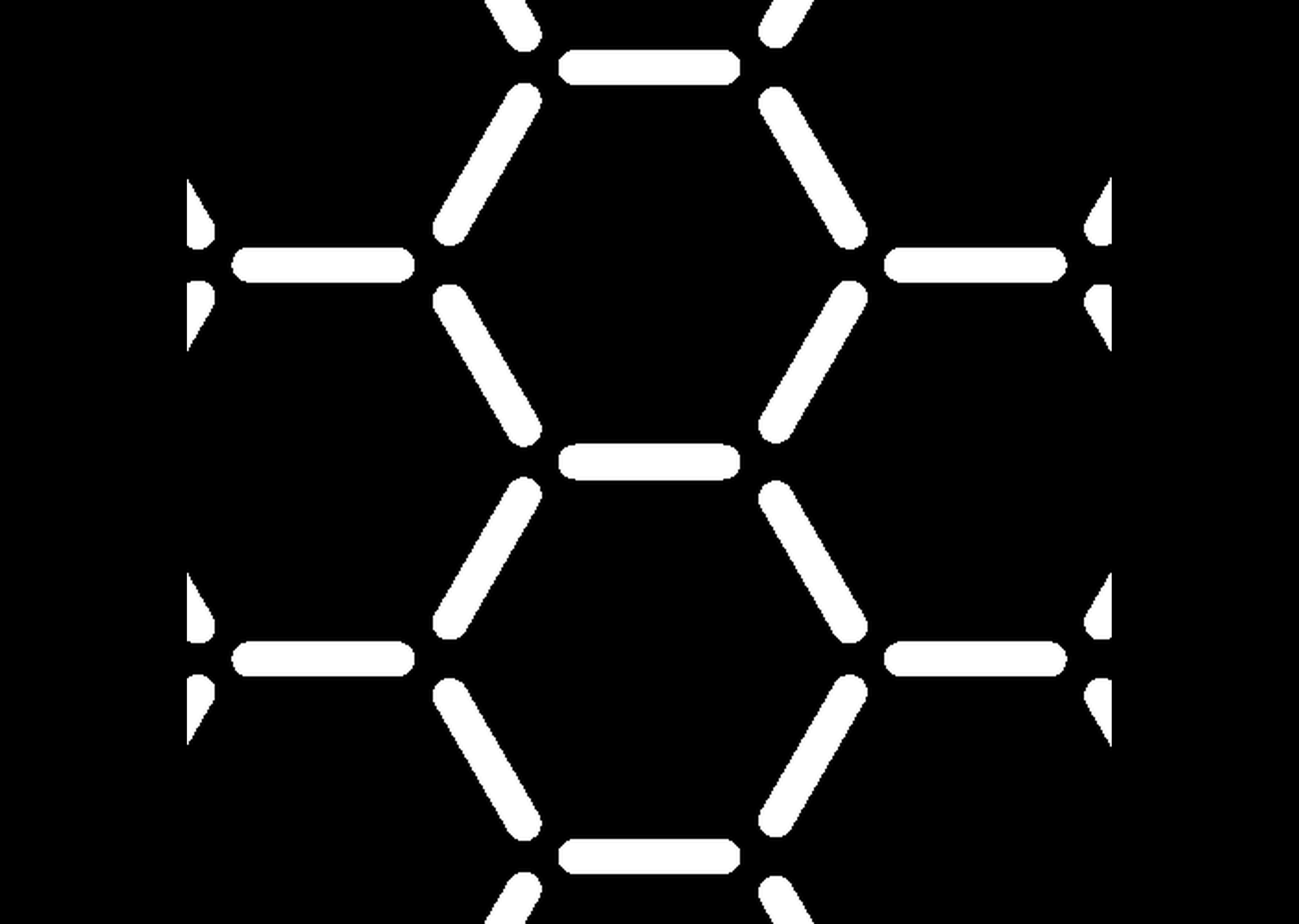} 
            & \includegraphics[width=0.2\linewidth]{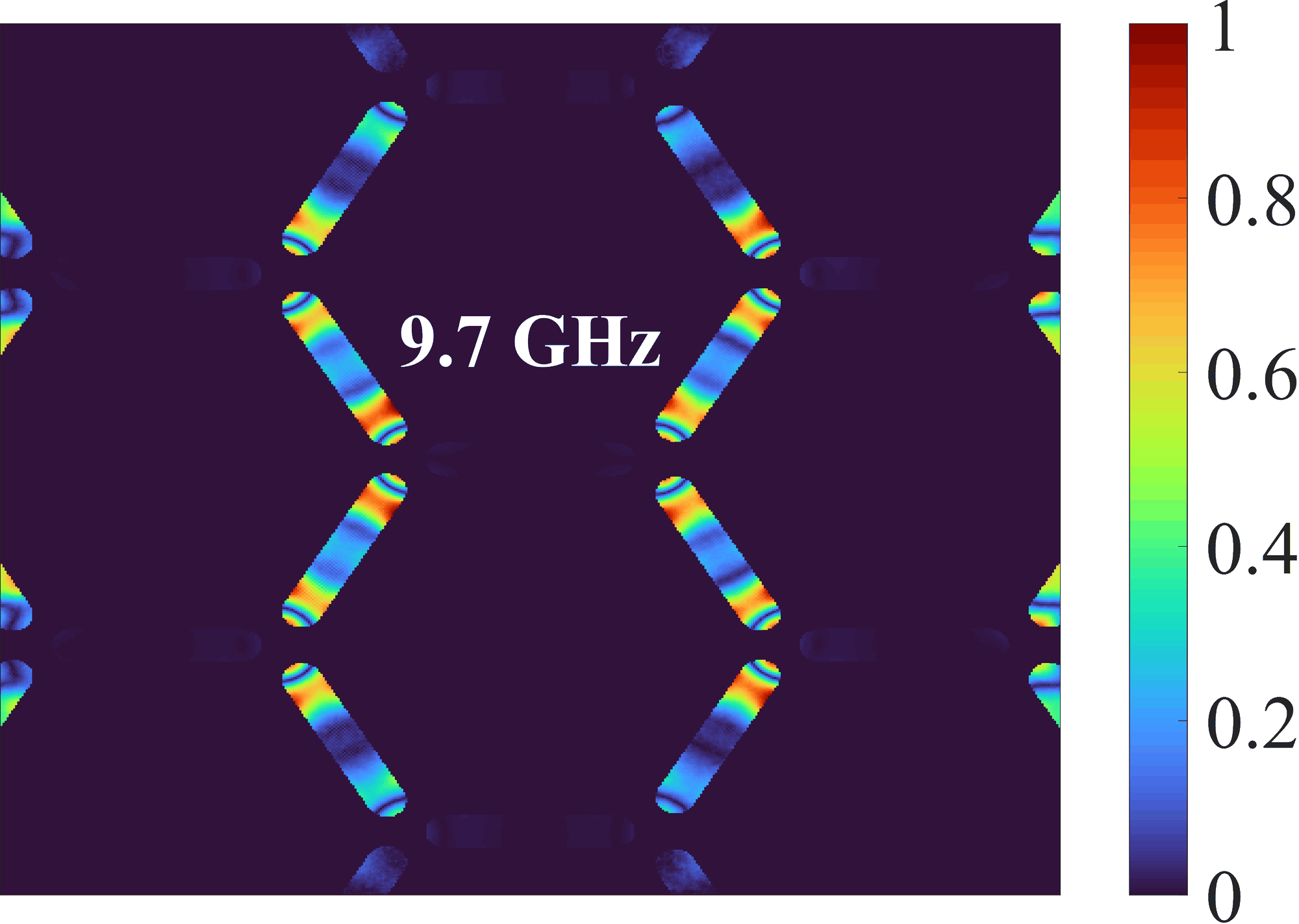} 
            & \includegraphics[width=0.2\linewidth]{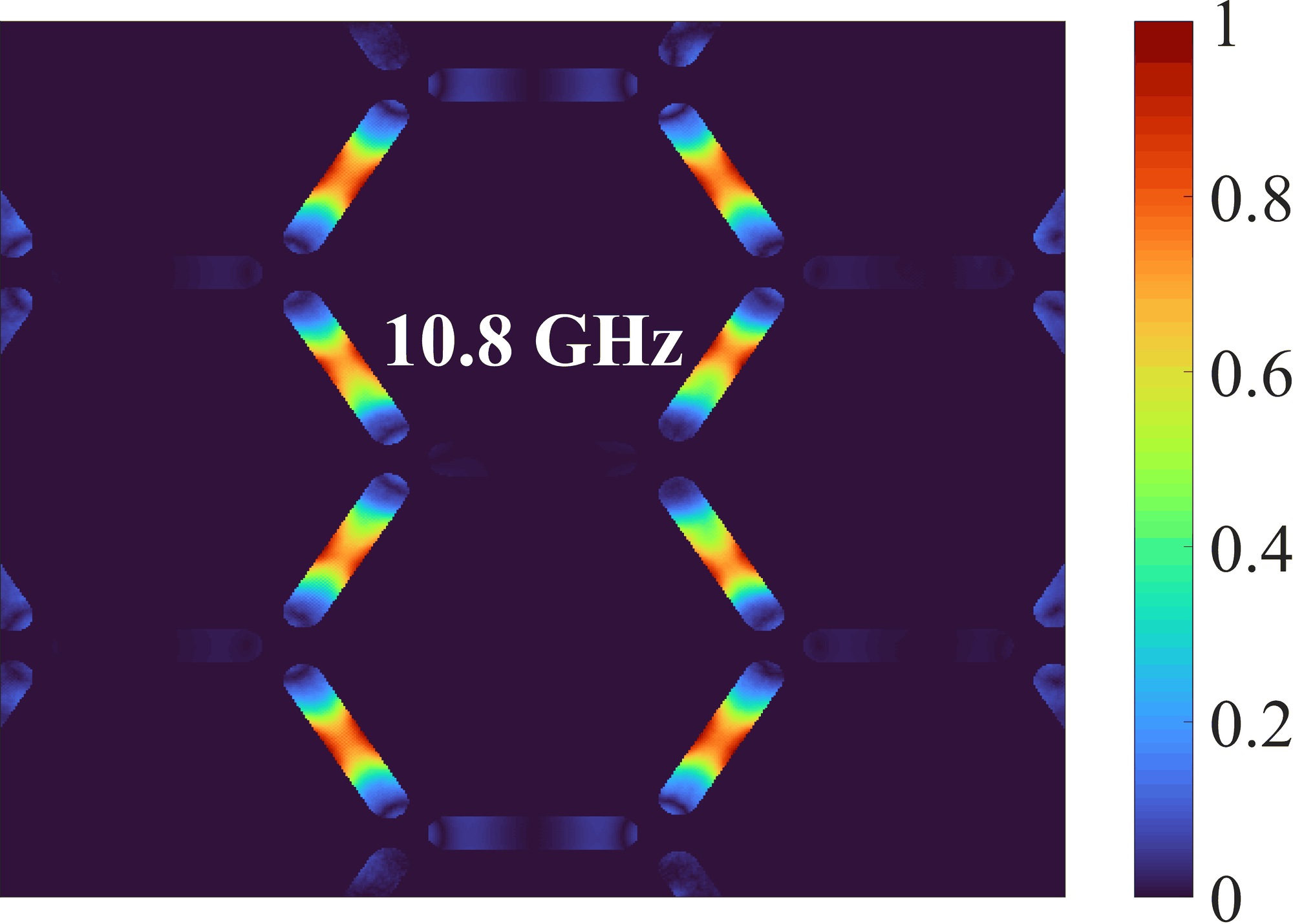} \\
        \hline
\centering       \raisebox{5\height}{80\,nm}  & \includegraphics[width=0.195\linewidth]{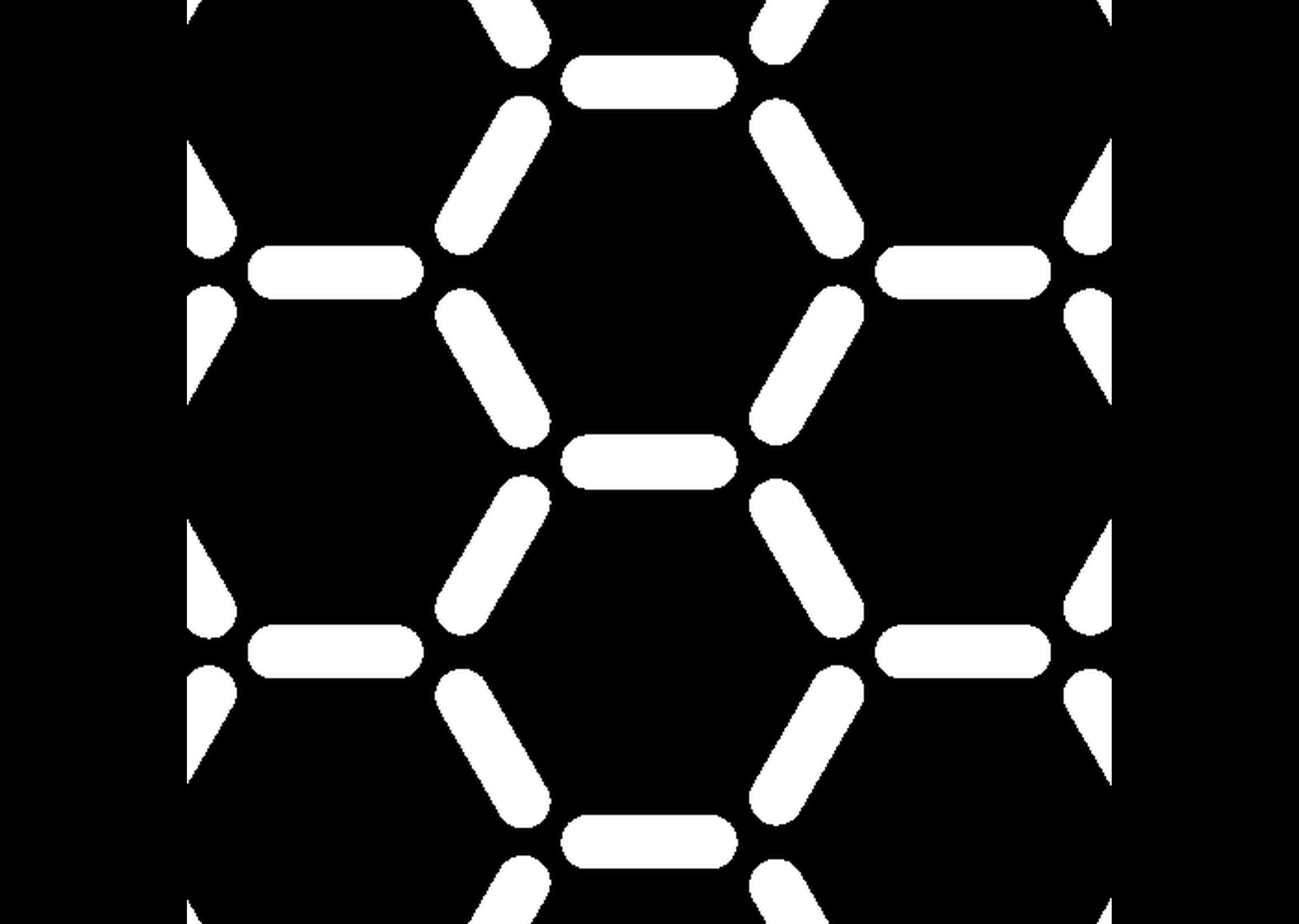} 
            & \includegraphics[width=0.2\linewidth]{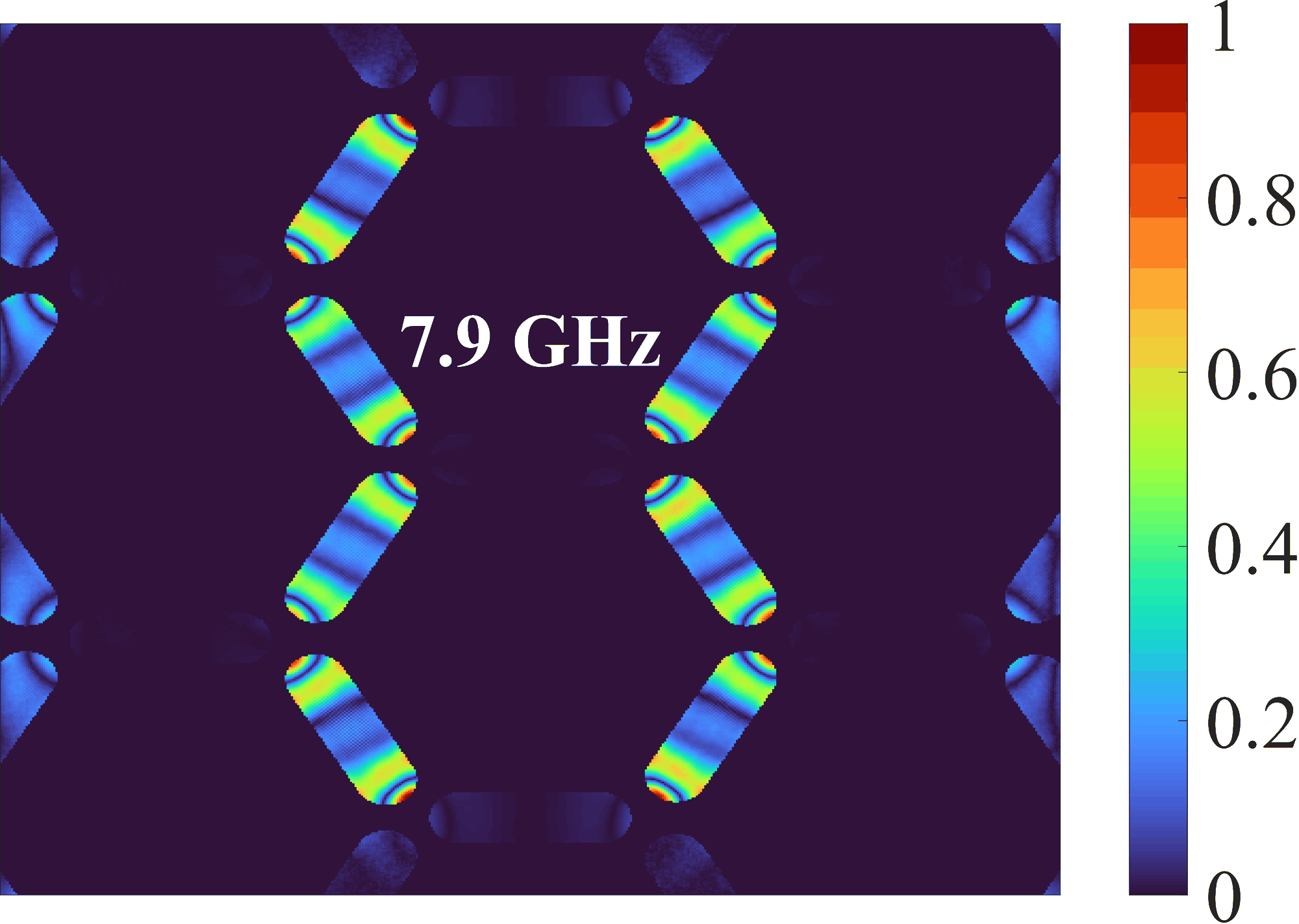} 
            & \includegraphics[width=0.2\linewidth]{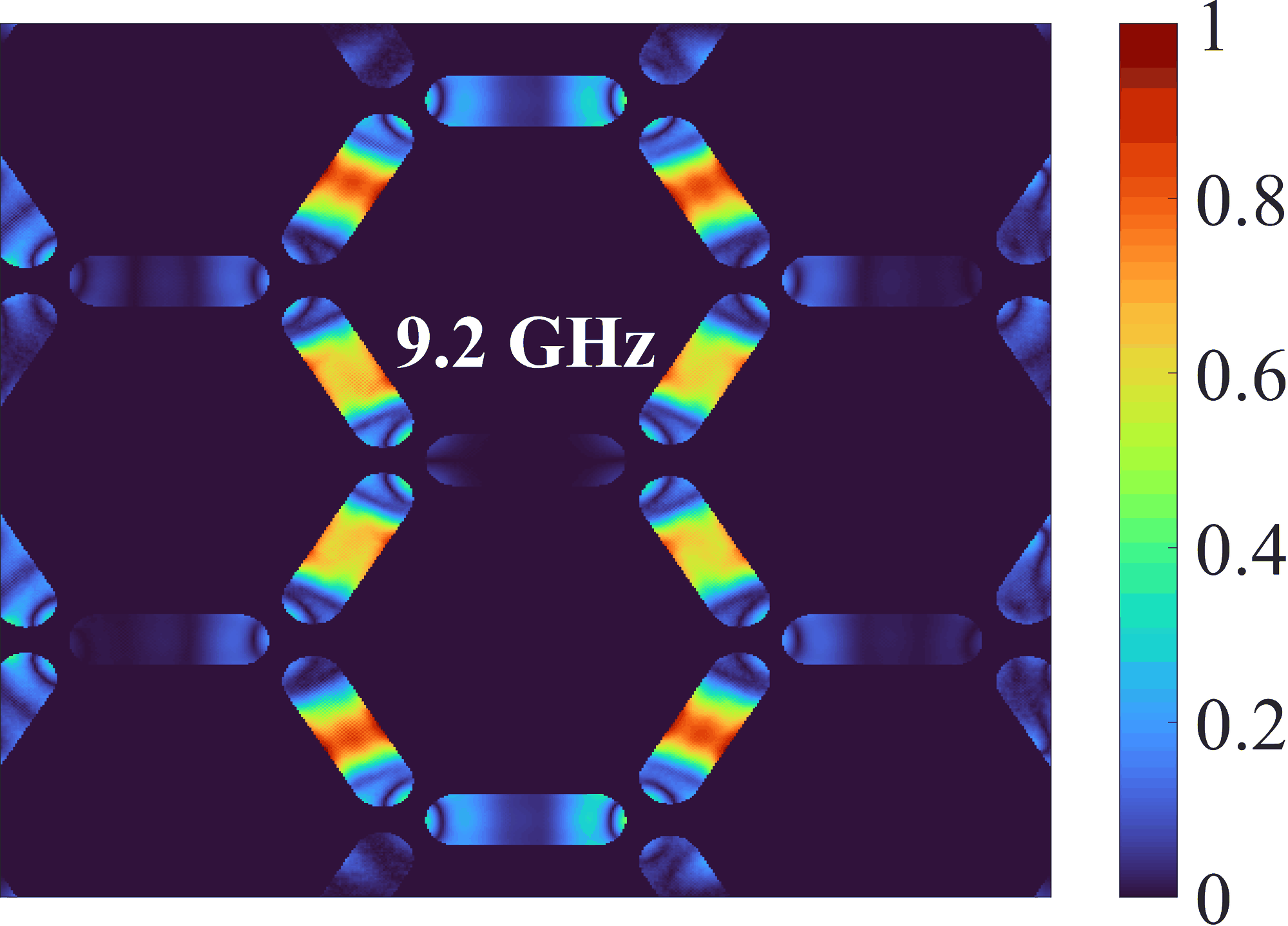} \\
        \hline
       \centering \raisebox{5\height}{110\,nm}  & \includegraphics[width=0.195\linewidth]{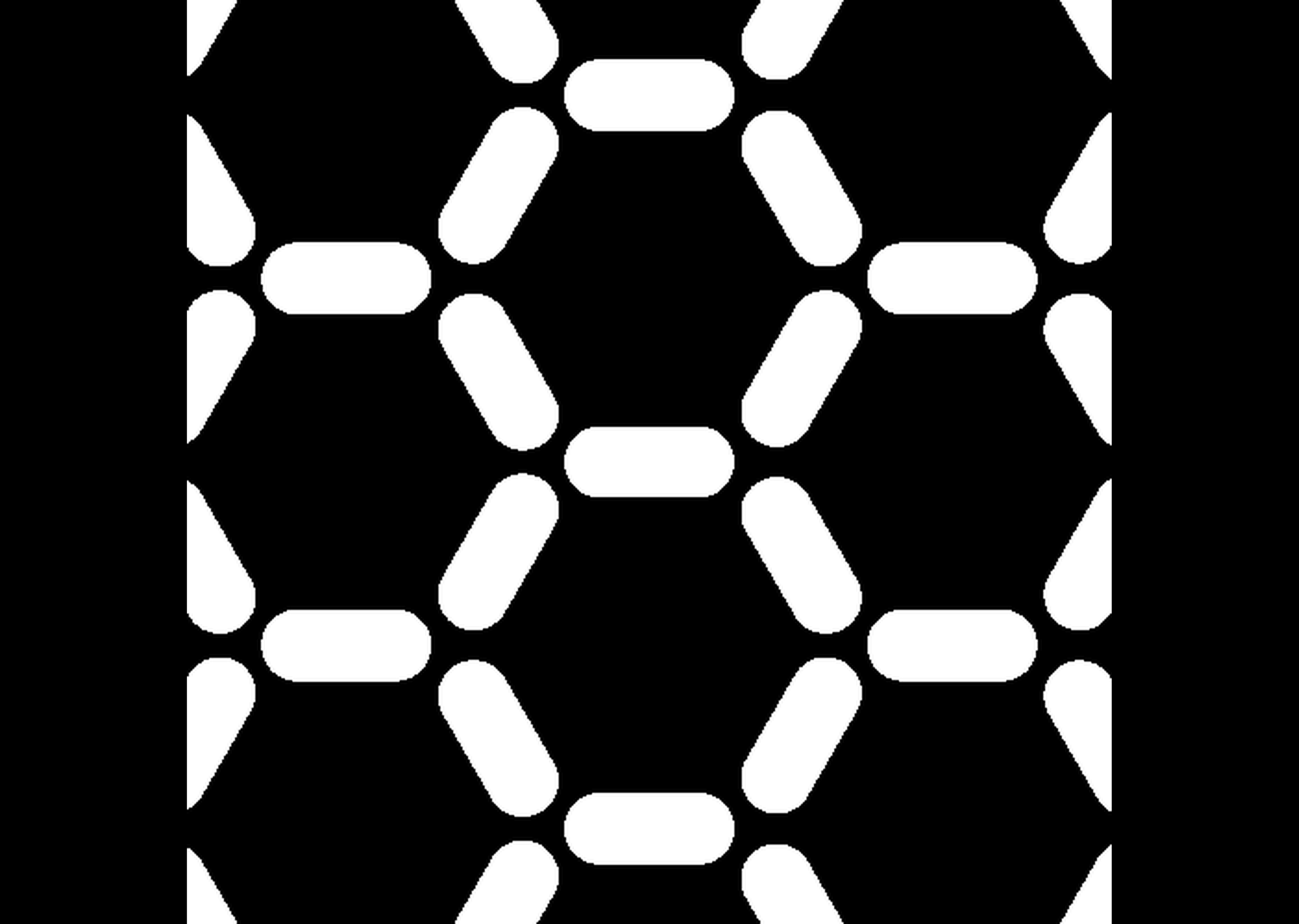} 
             & \includegraphics[width=0.2\linewidth]{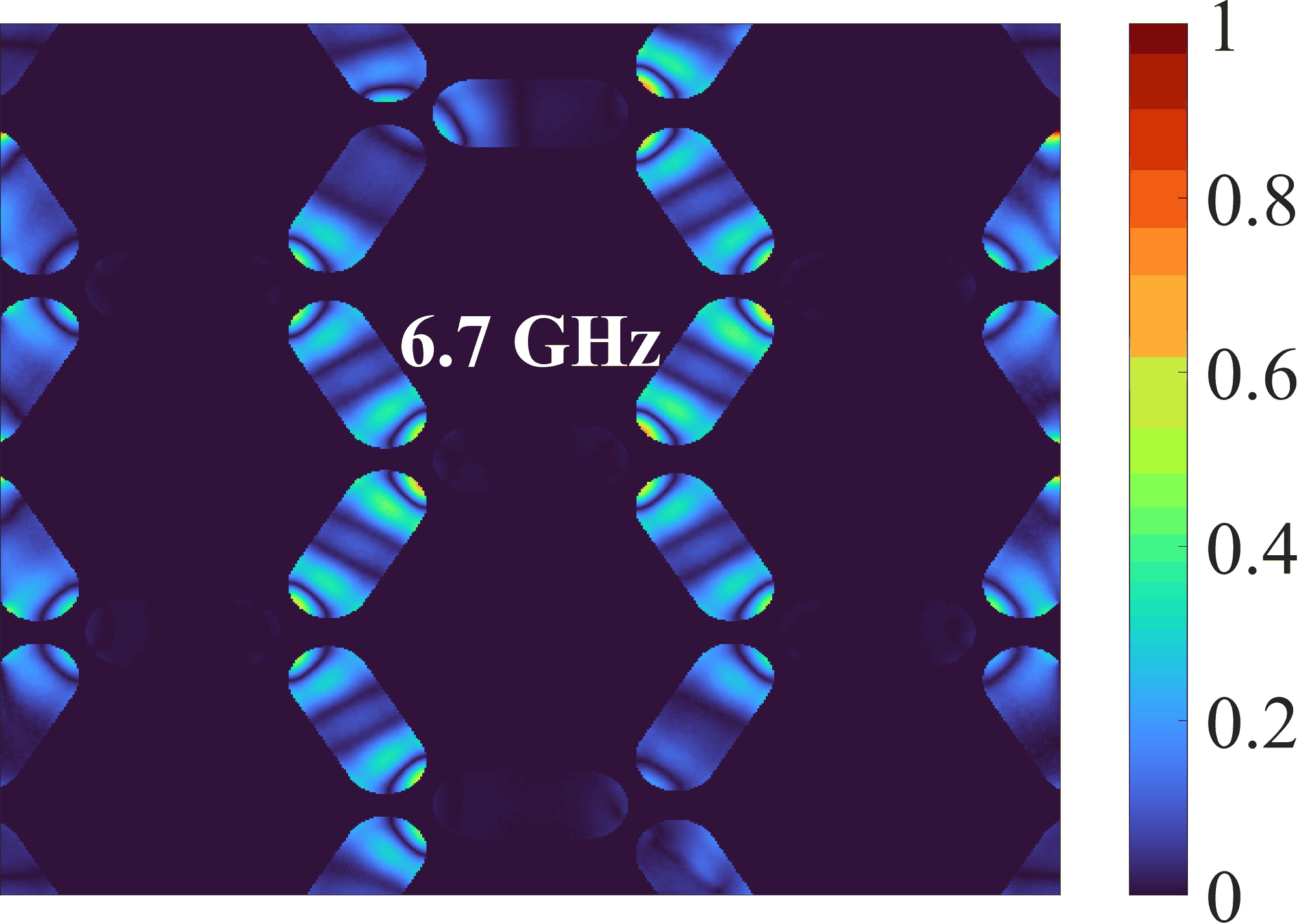} 
             & \includegraphics[width=0.2\linewidth]{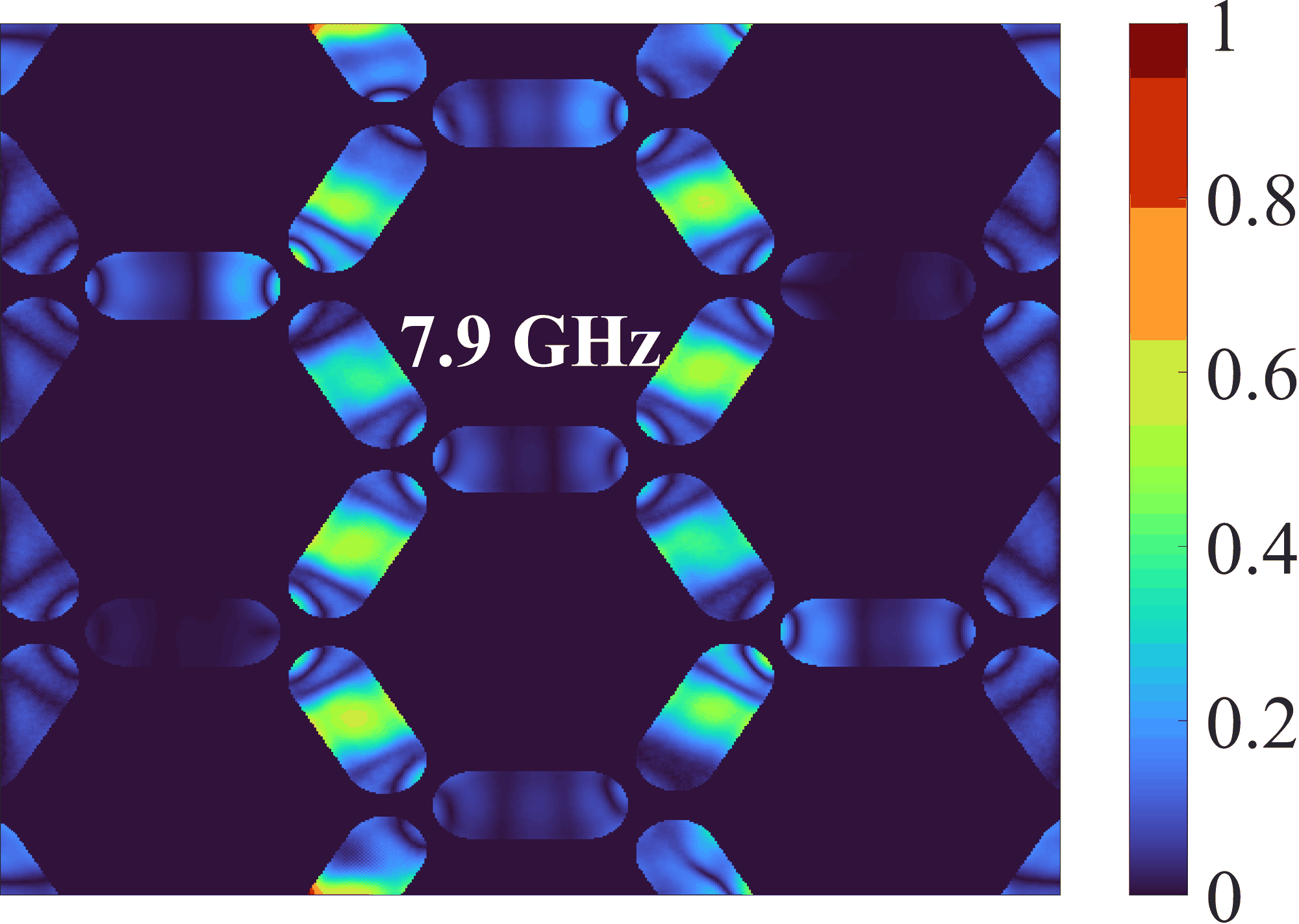} \\
        \hline
    \end{tabular}

    \caption{Top: (a) Hysteresis loops for nanoislands of widths = 50, 80, 110 nm and fixed length = 260 nm   (b) Resonance spectra for nanoislands of widths w = 50, 80, 110 nm and fixed length l = 260 nm
     Bottom: Comparative geometry schematics and resonance results for the three widths.  Annotated mode x, in the resonance spectra appear at frequencies 9.7 GHz, 7.9 and 6.76 GHz while annotated mode y appear at 10.8 GHz, 9.2 GHz and 7.9 GHz for the width of 50, 80, and 110nm respectively.}
    \label{tab1ab}
\end{table*}

\begin{table*}[!htph]   
   \begin{tabular}{|c|c|c|c|} 
        \hline
        Nano island & Geometry & Synthetic MFM & Final state \\
        \hline
\raisebox{7\height}{Wide 110\,nm}          &  
       \includegraphics[width=.19\linewidth]{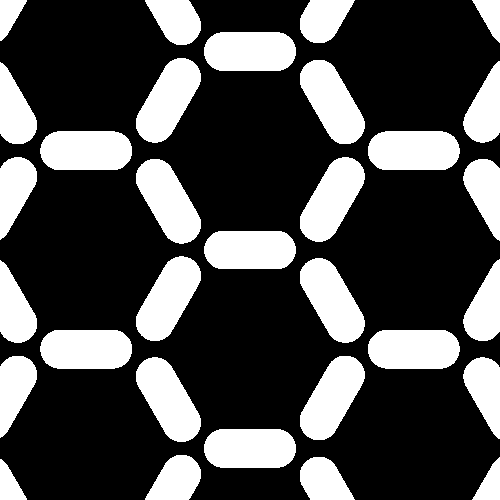} & 
       \includegraphics[width=.19\linewidth]{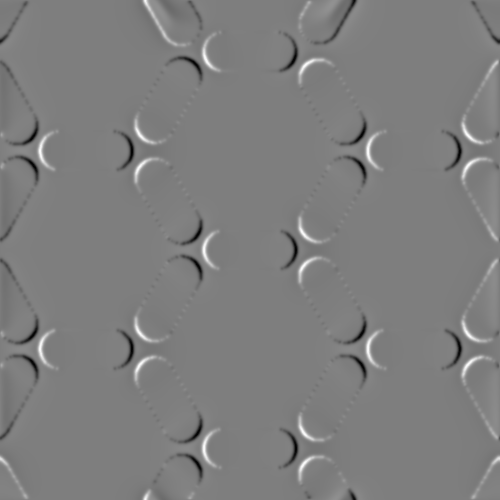} & 
        \includegraphics[width=.19\linewidth]{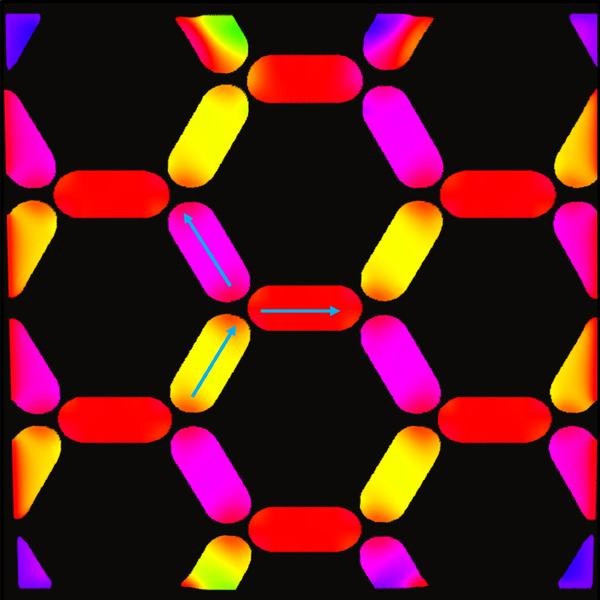} \\
      \hline
\raisebox{7\height}{Thin  60\,nm}          &  
        \includegraphics[width=.19\linewidth]{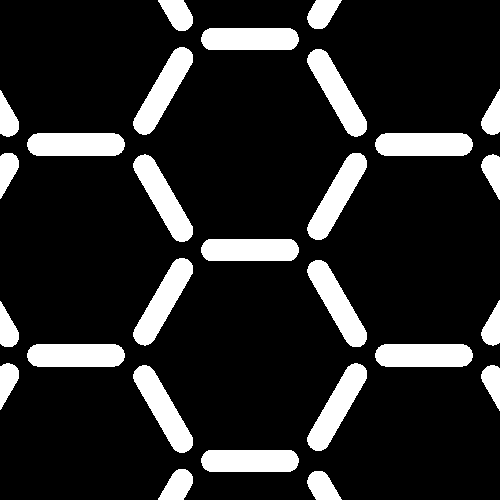} & 
        \includegraphics[width=.19\linewidth]{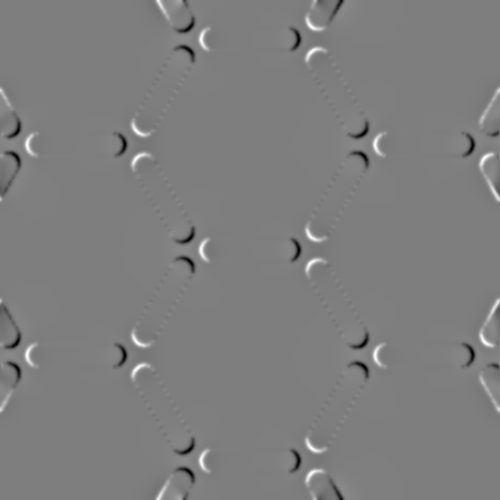} & 
        \includegraphics[width=.19\linewidth]{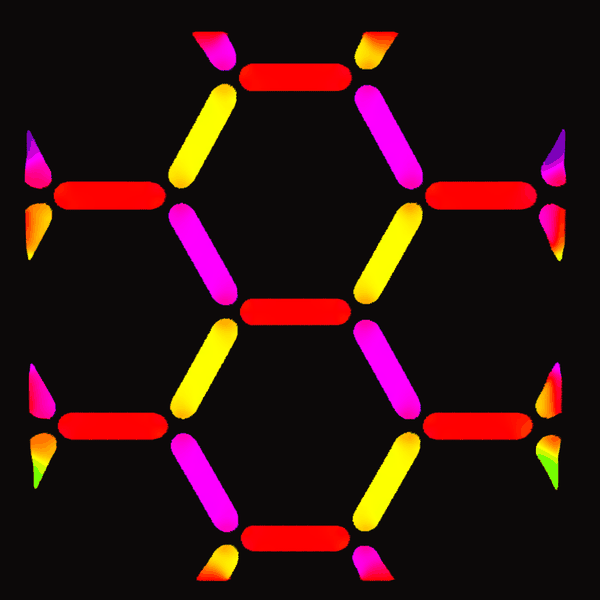} \\
        \hline
\raisebox{7\height}{Asymmetric Leg Length 310\,nm}          &  
        \includegraphics[width=.19\linewidth]{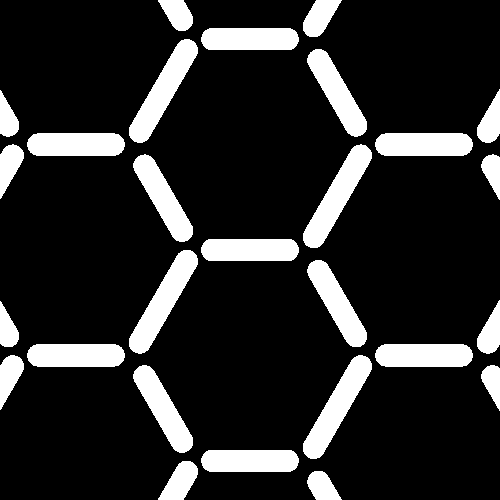} & 
        \includegraphics[width=.19\linewidth]{{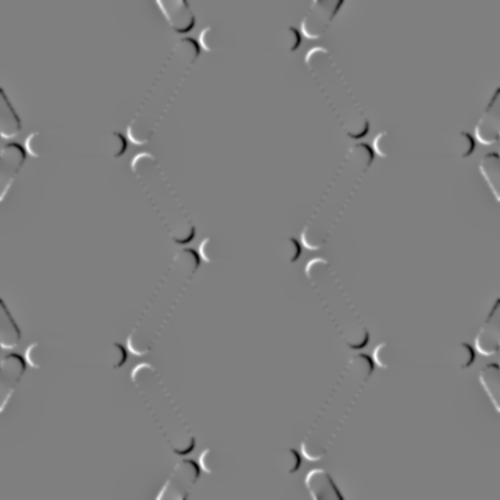}} & 
        \includegraphics[width=.19\linewidth]{{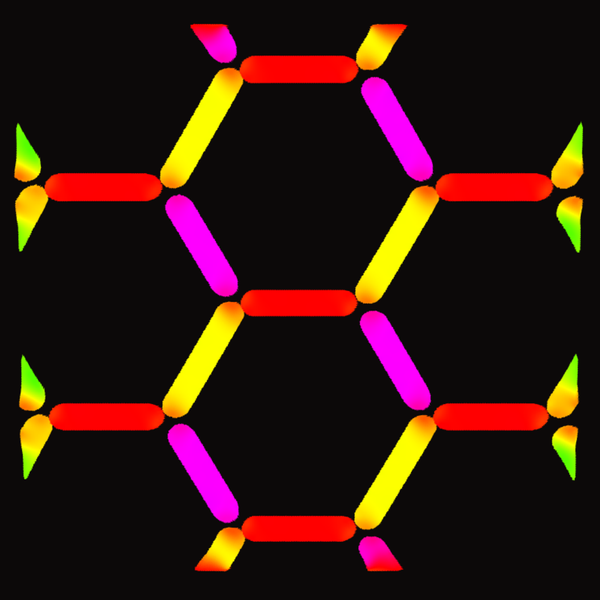}} \\
        \hline
   \end{tabular}
    
   \caption{geometry, magnetic force microscopy and equilibrium magnetization profile are shown. (MFM)-like images of relaxed vertex configurations for each geometry reveal more uniform and symmetric states in the wide ASI, moderately coupled behavior in the narrow case, and distorted vertex arrangements in the asymmetric geometry. For final state arrows are drawn to indicate the spin direction.}
    \label{tab1b}
\end{table*}

\begin{figure*}[!htph]
    \centering
\includegraphics[width=\linewidth]{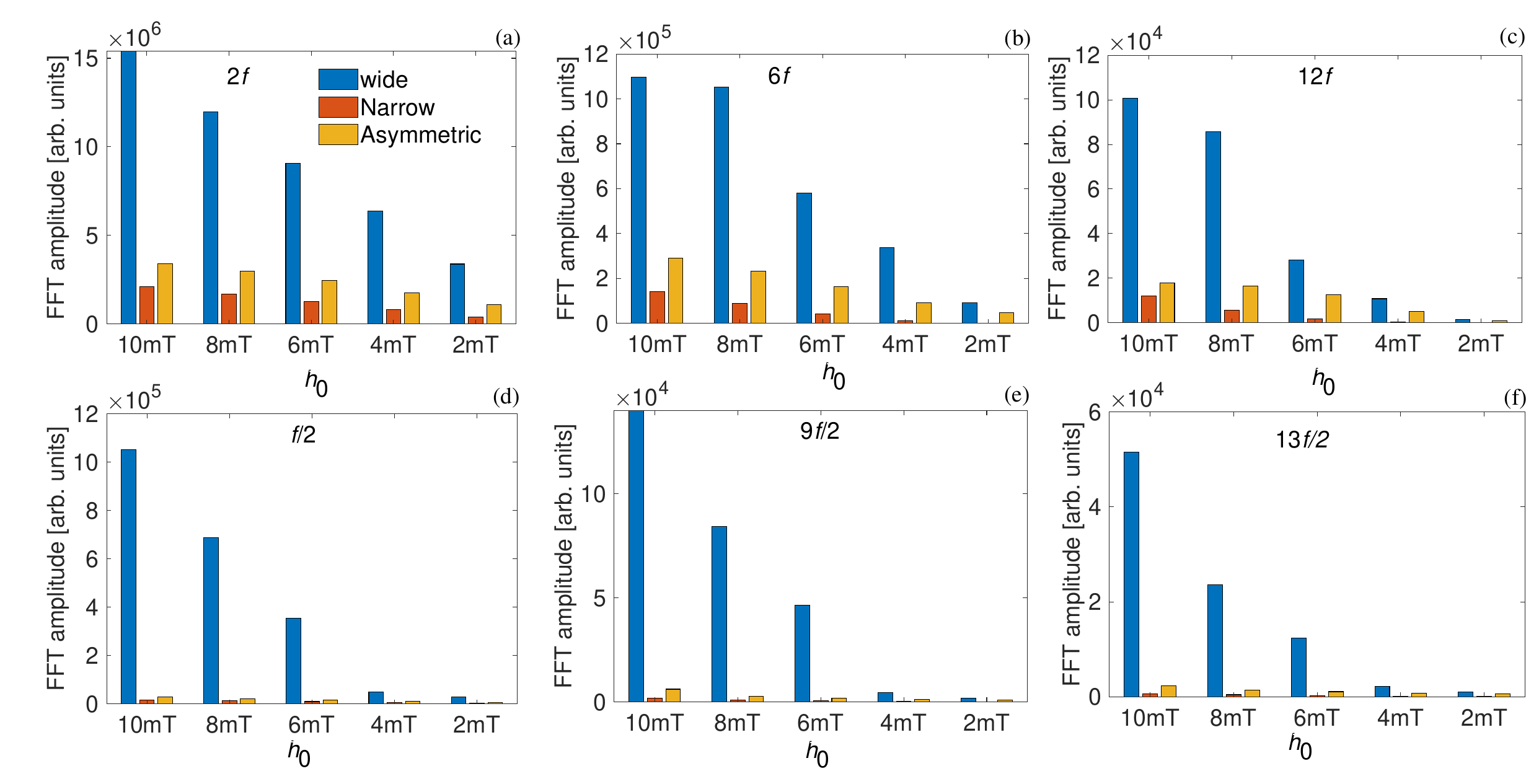}
\caption{
Harmonic and subharmonic amplitudes generated under microwave excitation for wide, narrow, and asymmetric kagome artificial spin ice. The wide geometry produces the strongest nonlinear response because its larger volume enhances dipolar interactions, whereas asymmetric leg length distorts the vertex configuration and suppresses coherent collective dynamics.}
    \label{fig2}
\end{figure*}

  

 \begin{figure*}[!htph] 
\centering
    \begin{tabular}{|c|c|c|c|}
        \hline
        Mode & 50 nm & 80 nm & 110 nm \\
        \hline
       \raisebox{7\height}{Mode X}  & \includegraphics[width=0.25\linewidth,height=.15\linewidth]{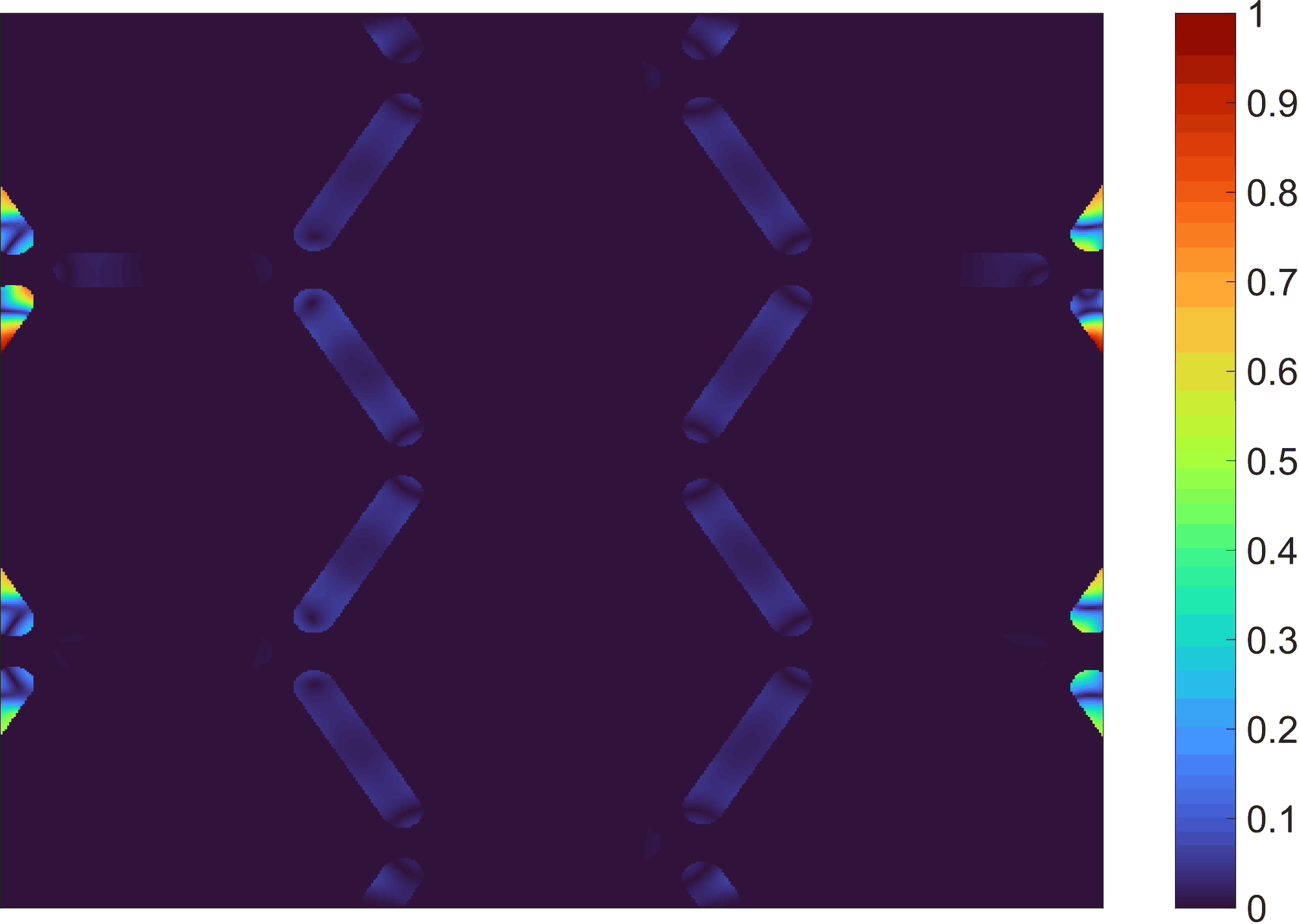}  
            & \includegraphics[width=0.25\linewidth,height=.15\linewidth]{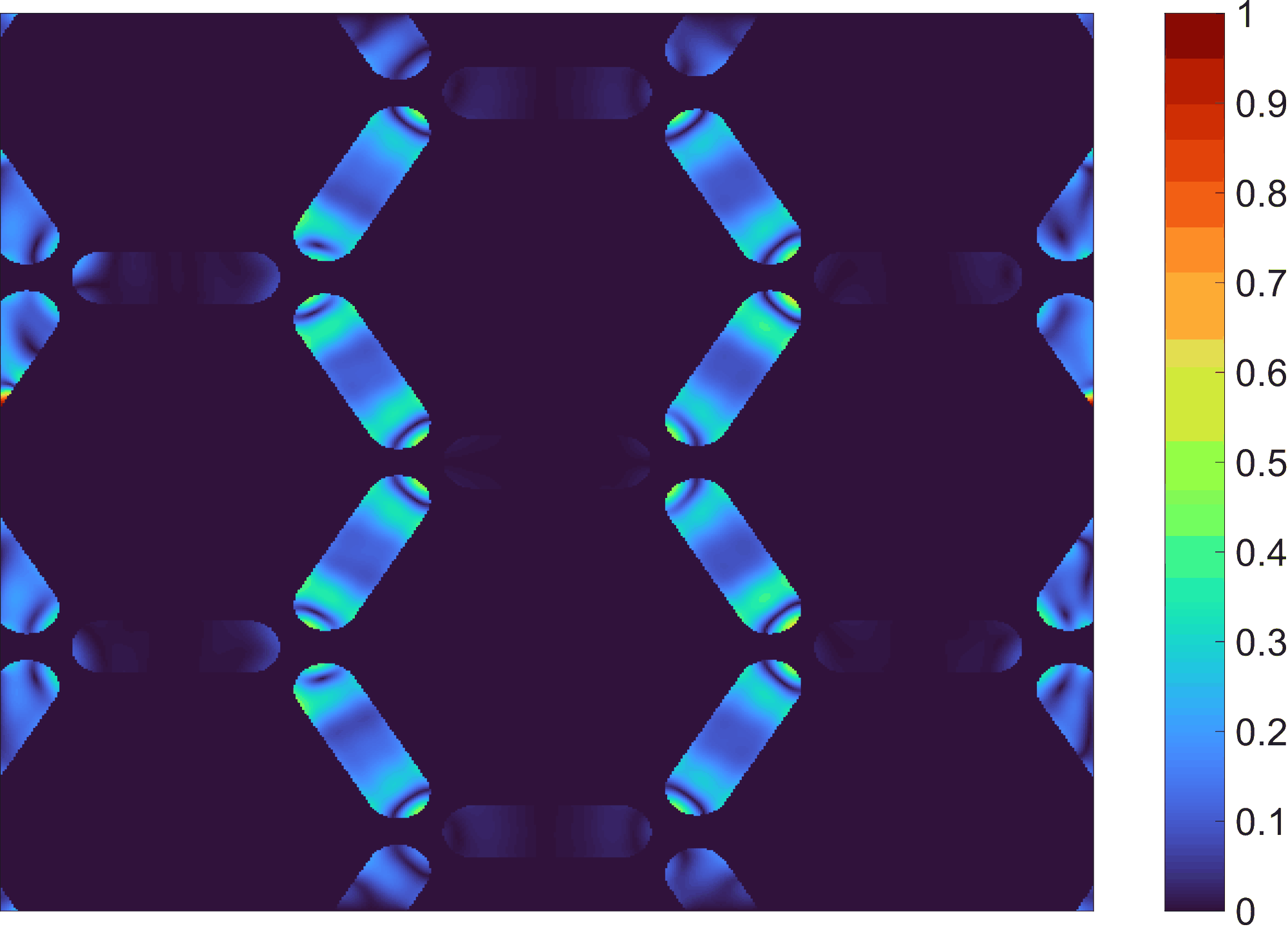} 
            & \includegraphics[width=0.25\linewidth,height=.15\linewidth]{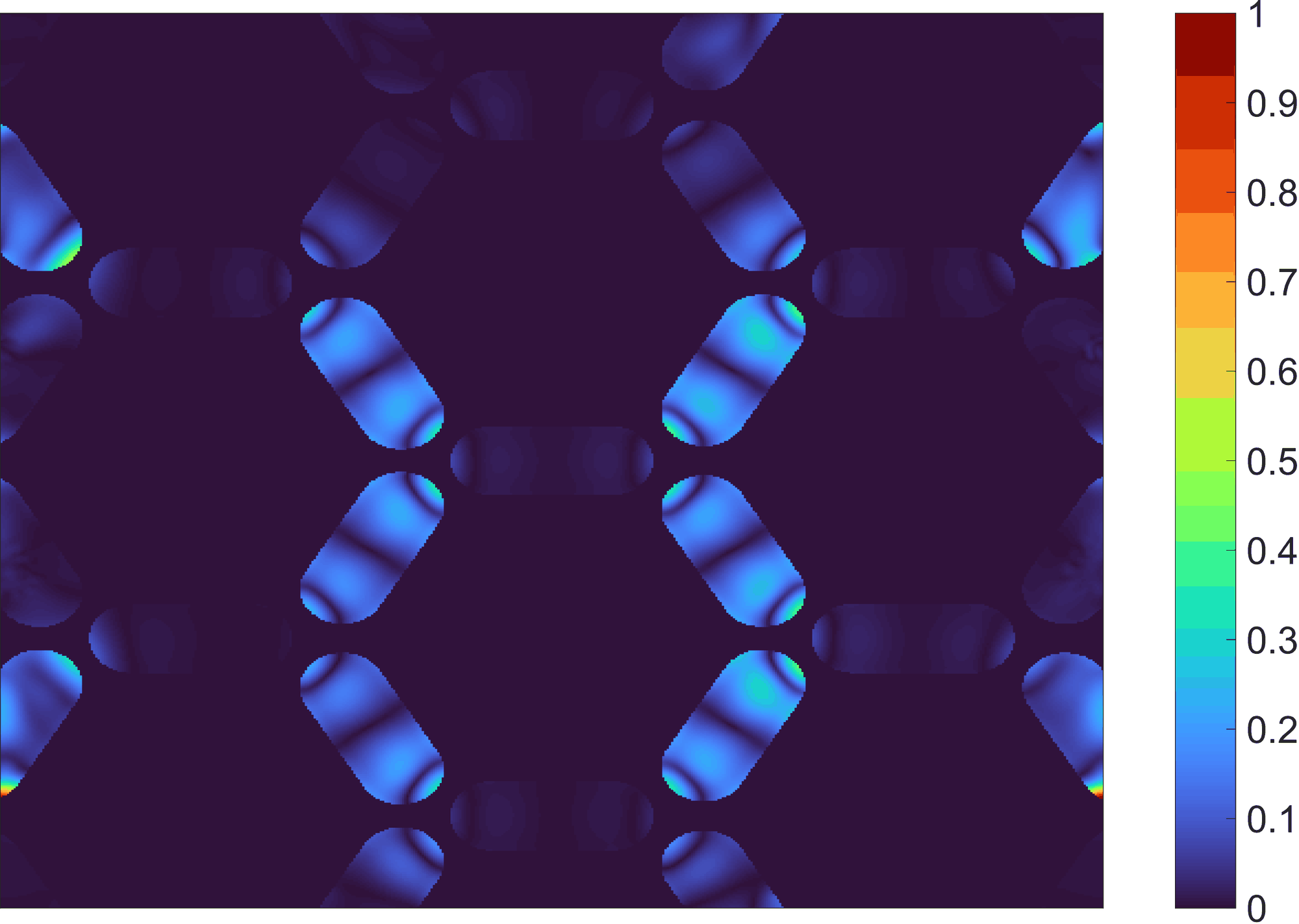} \\
        \hline
    \end{tabular}
  \caption{Parametric excitation of mode X for kagome nano islands with widths 50, 80 and 110\,nm (length = 260\,nm).  The mode red-shifts with increasing width ($f_X=9.7,\;7.9,\;6.76$\,GHz for 50, 80 and 110\,nm) while the applied field strength  decreases ($h_{0}=75,\;40,\;30$\,mT). The 110\,nm islands thus reach the largest, most spatially coherent amplitude at the smallest drive, indicating that island width both tunes the eigen-frequency and lowers the parametric threshold for nonlinear amplification.}
    \label{figaa}
\end{figure*}

\begin{table*}
    \centering
 
    \begin{tabular}{|c|c|c|c|}
    
        \hline
         Radius & Geometry &  $\hat{x}B_{\text{demag}}  $ & $\hat{x}B_{\text{exch}}   $
         \\ 
        \hline
      \raisebox{8\height}{R=110\,nm}   & \includegraphics[width=0.2\linewidth]{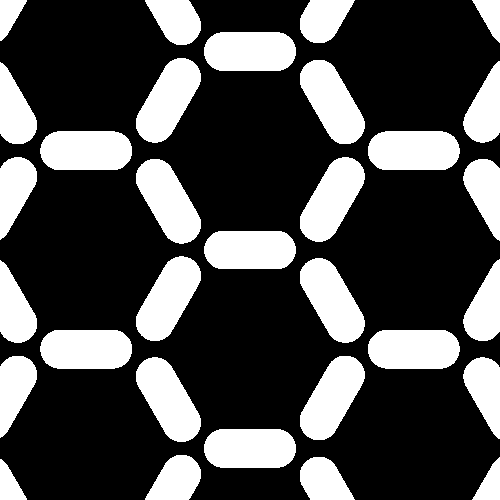}  &  \includegraphics[width=.25\linewidth,height=0.199\linewidth]{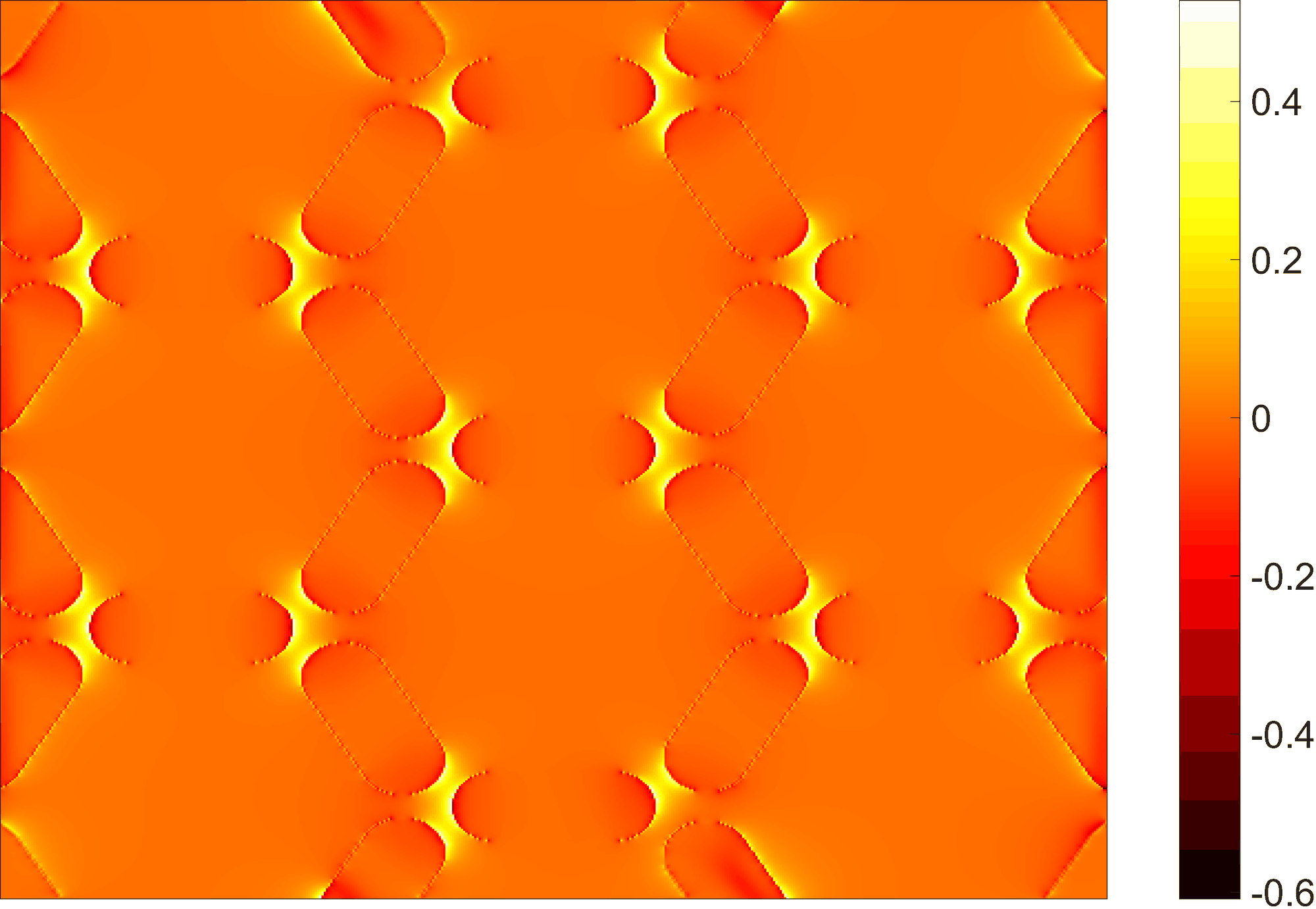}
  &\includegraphics[width=0.25\linewidth,height=0.2\linewidth]{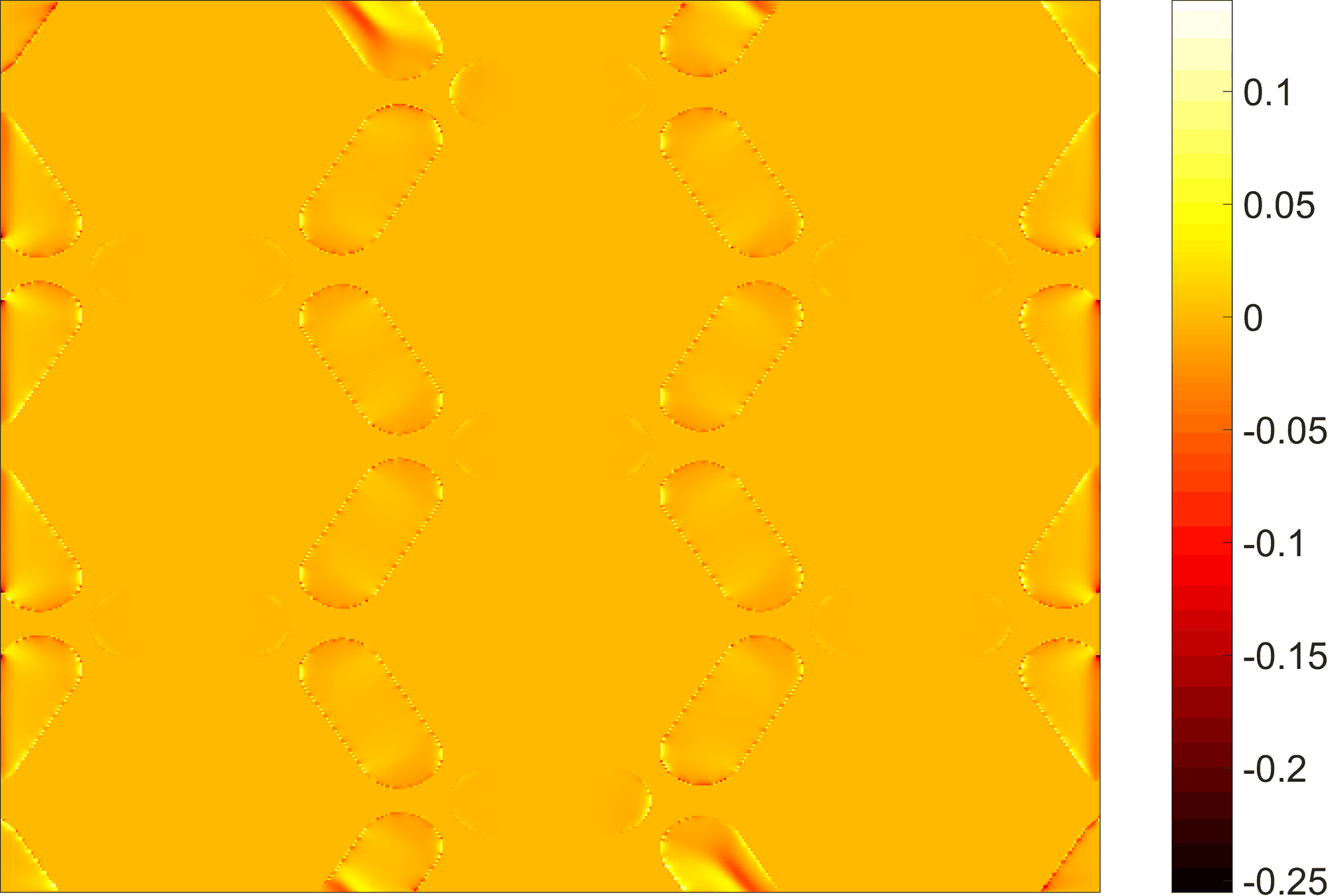} 
        \\
        \hline
       \raisebox{8\height}{R=120\,nm}  & \includegraphics[width=.2\linewidth]{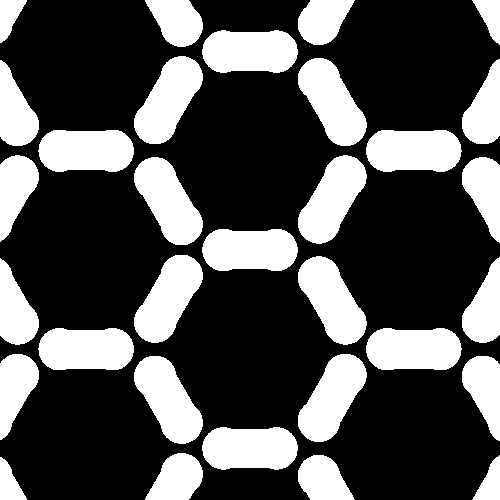}  &  
        \includegraphics[width=0.25\linewidth,height=0.2\linewidth]{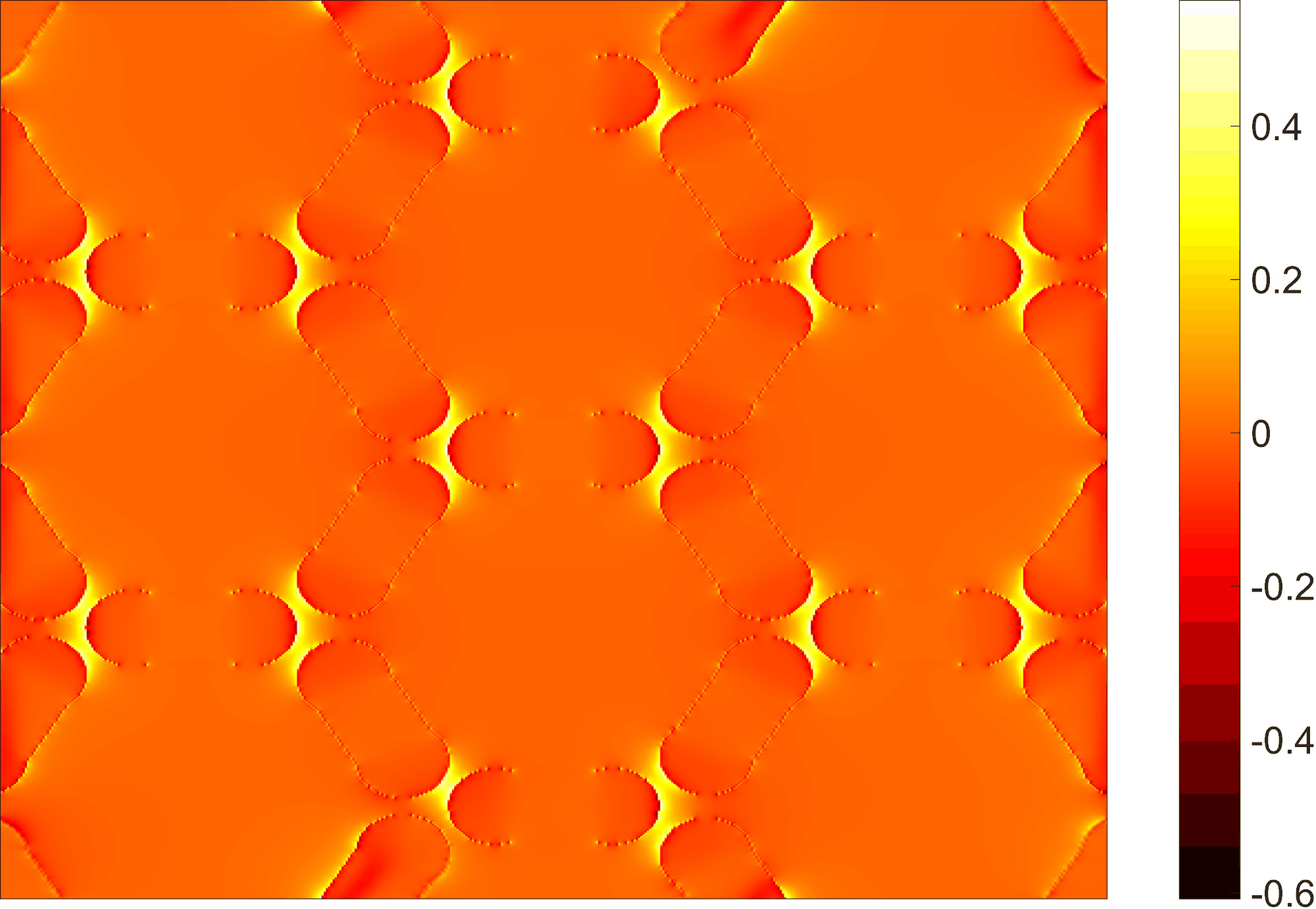} 
        & 
        \includegraphics[width=.25\linewidth,height=0.2\linewidth]{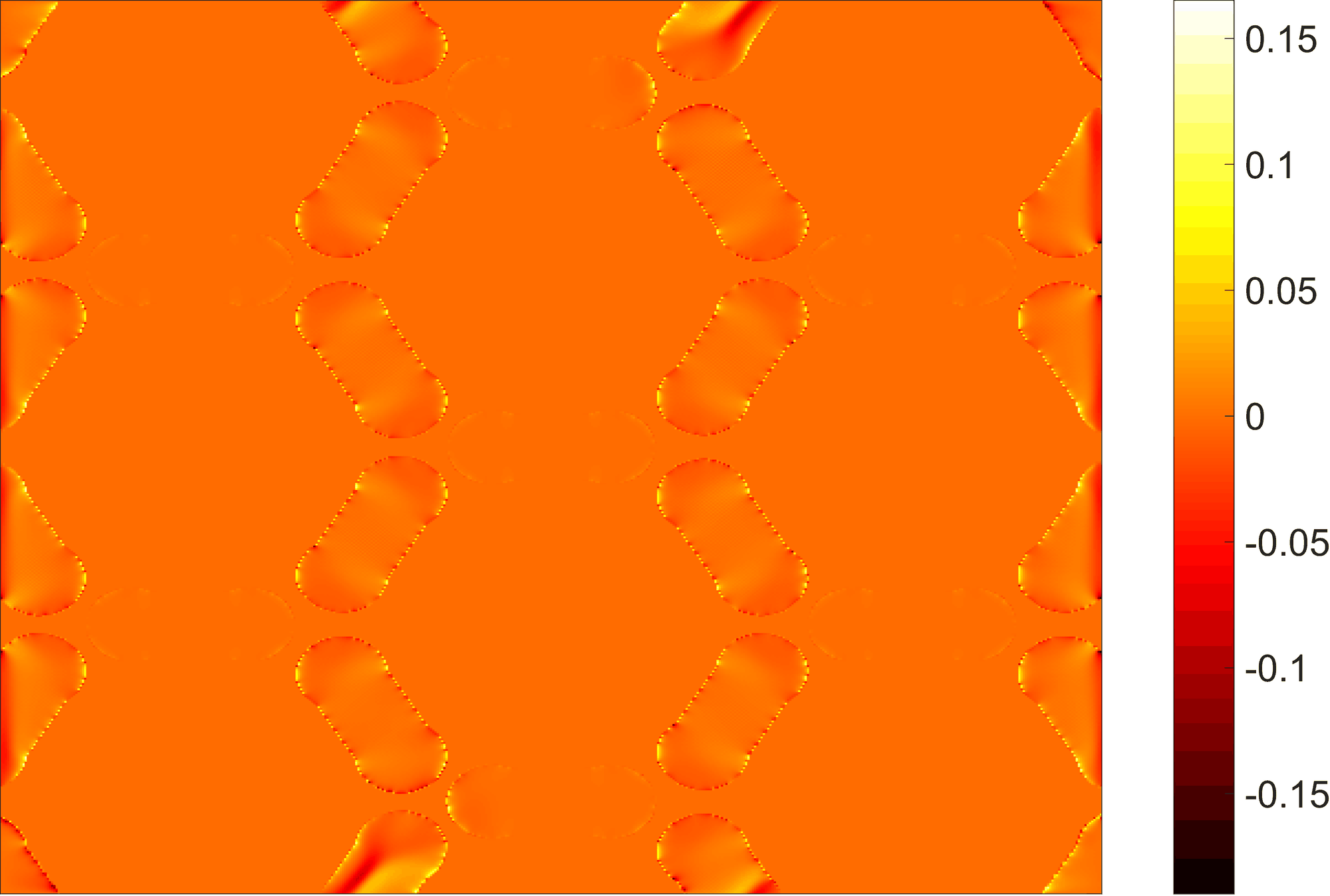}
        \\
        \hline
      \end{tabular}
    \caption{Impact of edge curvature on magnetization configuration and nonlinear spin-wave spectra in (ASI).  Two ASI geometries with nano island edge radii of 110 nm (ellipse-shaped) and 120 nm (dumbbell-shaped) are compared. The centre column shows the corresponding $X$ component of demagnetizing field distributions, with stronger and more localized fields observed near the curved ends of the dumbbell-shaped nano islands. Similarly, the right column shows $X$ component of exchange field distributions.
     }
    \label{tab2}
\end{table*}
\begin{figure}
    \centering
    \includegraphics[width=.9\linewidth]{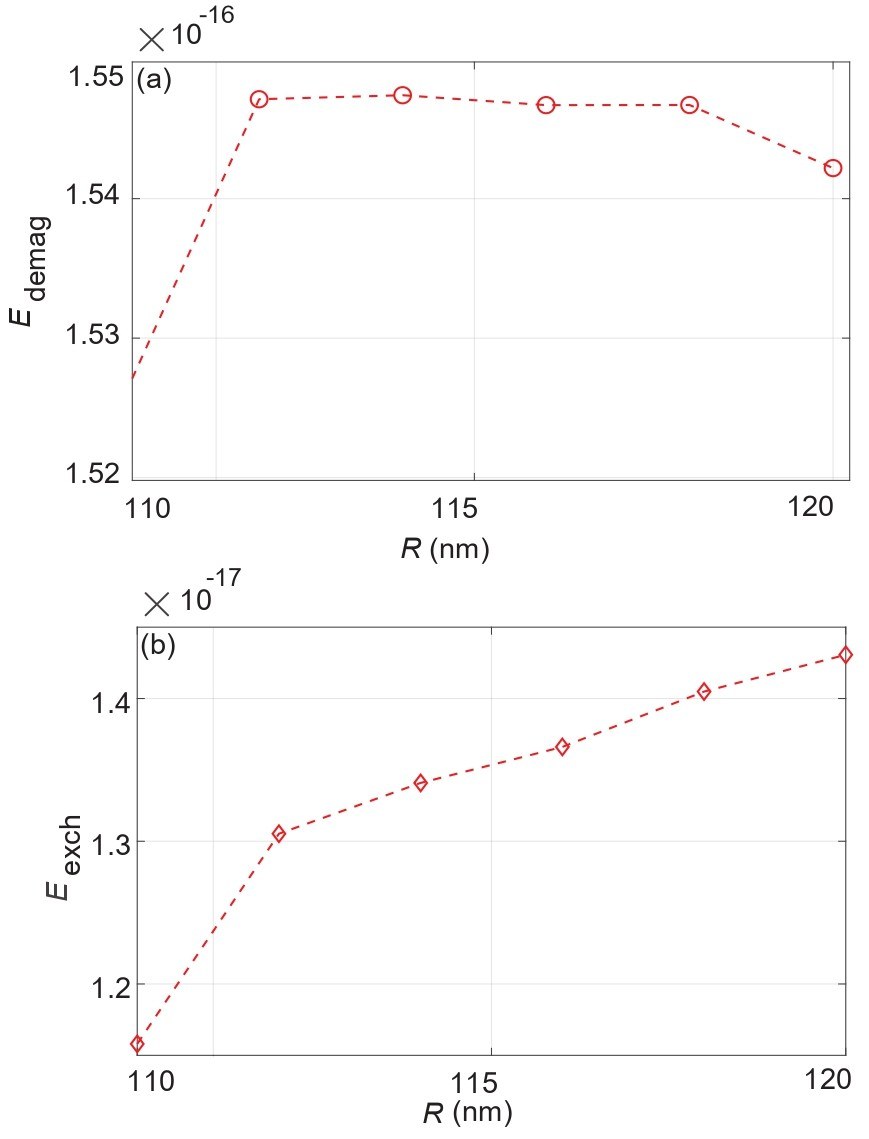}
    \caption{Effect of nanoisland edge-radius variation on the demagnetization and exchange energies is shown. A 1.16 \% energy difference in $E_{\text{demag}}$ and 23.6 \% energy difference in $E_{\text{exch}}$ is visible when edge radius changes from 0-10nm.}
    \label{fig_e}
\end{figure}




\begin{figure*}[!htph]

    \centering

    \includegraphics[width=\linewidth]{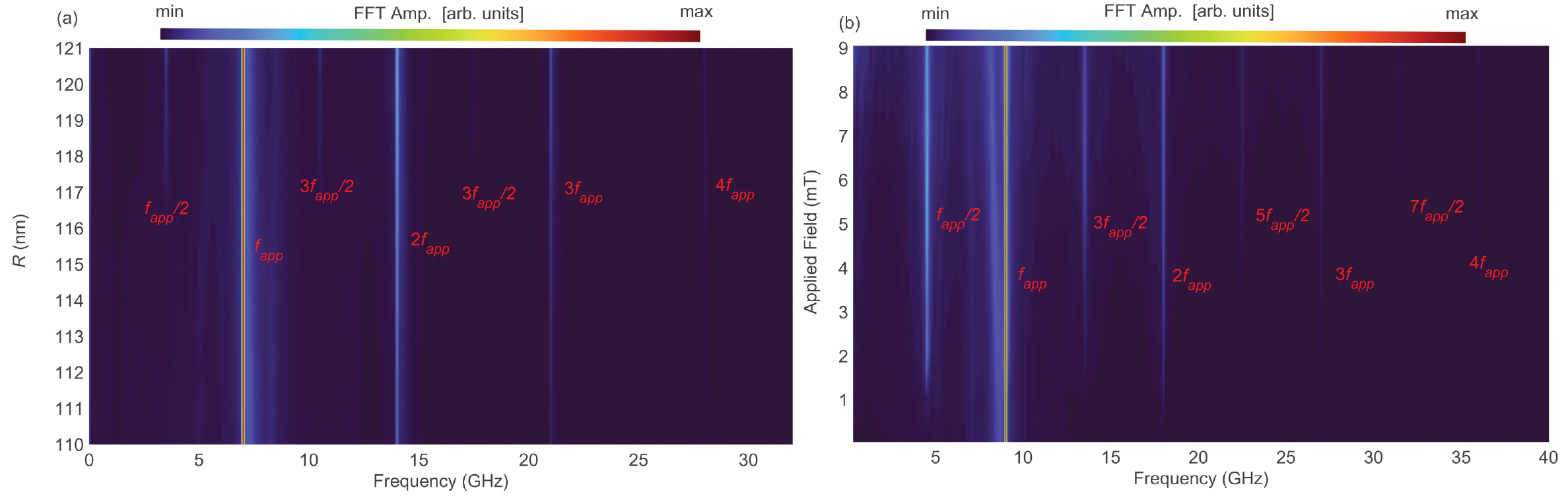}
    \caption{(a) Harmonic spectra and field maps for dumbbell shaped kagome artificial spin ice. Increasing the edge radius promotes subharmonic generation by localizing demagnetizing and exchange fields near the island ends, demonstrating that tip curvature can trigger nonlinear scattering without changing the island footprint. This graph was computed for an applied field strength of 2mT and frequency of 7 GHz. (b) Illustrated is the field–frequency map of the magnetization under in-plane excitation at an applied frequency of 9 GHz. }
    \label{fig4}
\end{figure*}

\begin{figure}[!htp]
	\centering
		\includegraphics[width=\linewidth]{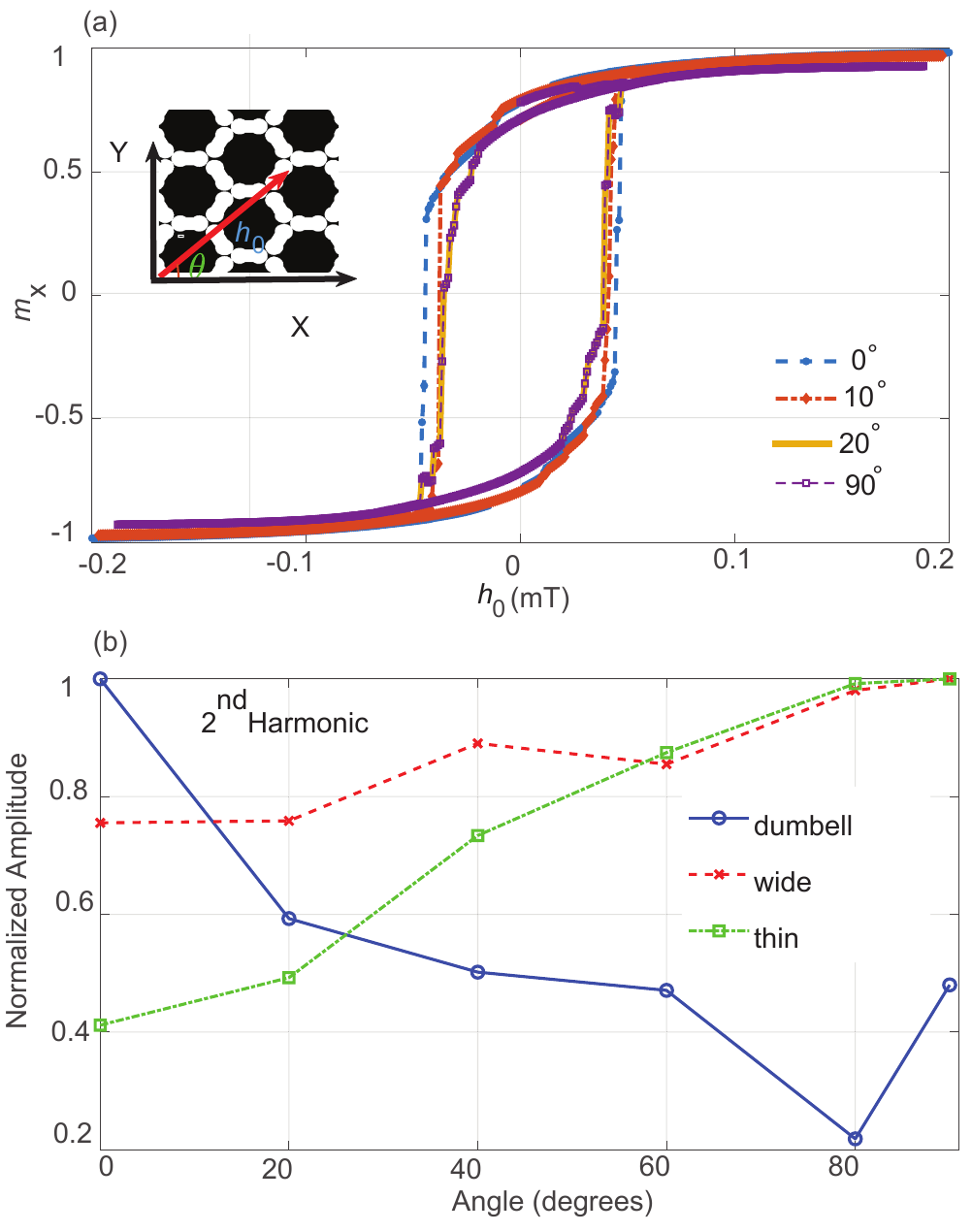}
	\caption{(a) Influence of the applied-field angle on hysteresis loops arising from nano island edge curvature. (b) Field‑orientation tuning of harmonic generation: normalized amplitudes of the 2nd harmonics under a 7 GHz, 10 mT excitation plotted as a function of the excitation angle.  Dumbbell shapes show maximum response at 0$^\circ$ (field parallel to symmetry axis), whereas wide/thin islands peak at 90$^\circ$ (perpendicular to easy axis), revealing geometry-dependent magnetization dynamics.  }
	\label{fig5}
\end{figure}


\section{Simulation Details}
We simulate the resonance spectrum of ASI by constructing the kagome structured nano islands. These nano islands have elongated shapes, resembling rounded rectangles or ovals that have been stretched as shown in Fig. \ref{fig1a}. The study systematically examines nano islands with widths ranging from $20<w<110$ nm and length from $260<l<310$ nm, keeping the thickness at a constant $15$ nm. For more specific geometric details, please refer to the codes \cite{aaa}. To simulate the  magnetization dynamics, we adopt MuMax3 that numerically solves  the Landau–Lifshitz–Gilbert (LLG) equation,
\begin{equation}
	\frac{ \partial \boldsymbol{m}}{ \partial t}=-\gamma \boldsymbol{m} \times \boldsymbol{H}_{{eff}}+\alpha \boldsymbol{m}\times \frac{ \partial \boldsymbol{m}}{ \partial t},
	\label{eq3}
\end{equation}
where $\boldsymbol{H}_{\mathrm{eff}} = -\delta E / \delta \boldsymbol{M}$ is the effective field. 
Here, $E$ is the magnetic energy density, $\boldsymbol{M}$ is the magnetization vector, $\gamma$ the (absolute) gyromagnetic ratio, $\alpha$ the Gilbert damping constant, and $\boldsymbol{m} = \boldsymbol{M}/M_s$ is the normalized magnetization, with $M_s$ being the saturation magnetization. 
In our simulations, the energy density includes exchange, demagnetization, and Zeeman contributions. 
 Throughout this paper, we have used Permalloy material (unless otherwise stated) to solve Eq. \eqref{eq3},  $M_s=800$ kA/m,  $A=13$ pJ/m,   and $\alpha=0.01$ with the cell size being $2.5\times2.5\times1\text{ nm}^3$ and the sample size $1250\times1250\times15\text{ nm}^3$.

\section{Results}
A major advantage of ASI lies in its ability to support a wide range of geometric configurations. The properties of ASI can be configured not only by using different geometries but also by changing the aspect ratio of the similar geometries. Although the aspect ratio is altered, it is beneficial to ensure that each nano island maintains macrospin behavior. In ASI, hysteresis loop simulations are commonly performed to verify that each nano-island behaves approximately as a macrospin. Magnetic properties such as coercivity and remanence can be highlighted by the study of hysteresis loop simulation.  Therefore, in Figs. \ref{fig1} (a) and (b), the hysteresis loop and the effect of the aspect ratio on the resonance spectra are illustrated. An increase in the width of the nano islands from 50 to 110 nm appears to result in narrowing of the hysteresis loop. This result suggests that the narrow nano islands exhibit high coercivity while the wide ones have lower coercivity, indicating the potential for high dipolar coupling for wide nano islands.

The spin wave resonance spectrum is obtained by perturbing the equilibrium magnetization of the kagome ASI with a microwave magnetic field \({h}(t) = h_{0} \, \text{sinc}[2 \pi f_{c} (t-t_{0})] \hat{x}\) applied over a period \(t\), where \(h_{0}\) is the field strength and \(f_{c}\) is the cutoff frequency. The initial magnetization of the nano-islands is $\bold{m}=(1,0,0)$, and the ground state was determined without applying any external bias field. To induce resonance, nano islands were excited under $f_c = $ 50 GHz and of $h_{0}$ = 20 mT with an offset of $t_0=1$ ns. The simulation was run for 10 ns and data was recorded after every 9.1 ps. The microwave excitation generates standing spin waves, with resonance modes localized at the edges of the nano islands as well as a distinct resonance mode within their interiors. These modes are referred to as edge or bulk modes, respectively. In Fig. \ref{fig1} (b), the responses of the kagome ASI with multiple widths are illustrated. The spectra indicates that the variation of the width of nano islands changes the resonance spectra. The resonance peaks shift to the lower frequencies suggesting a decrease in effective field as a result of an increase in the dipolar field by the increase in the volume of kagome ASI. Table \ref{tab1ab}, illustrates the corresponding geometries and their resulting modes X and Y. Notice that these modes appear at different frequencies by changing the width of nano islands.

In a non-connected system the dipolar fields can be the dominant source of nonlinearity due to its long-range interactions between the magnetic moments as oppose to  exchange fields. We are now going to investigate nonlinear scattering in kagome (ASI). For this purpose, we used three different types of nano islands, i.e., thin, wide, and asymmetric. The asymmetry in the kagome ASI is achieved by increasing the length of one of the nano islands. This allows to tailor the vertex degeneracy of the ground state by controlling the competing dipolar interactions between the neighboring nanomagnets. The corresponding geometries, synthetic magnetic force microscopy (MFM) computing highlighting vertex charge distribution and island polarity orientation, and relaxed state are shown in table \ref{tab1b}.  From the synthetic MFM it is evident that the wide geometry exhibits more distinct and symmetric vertex states, indicative of stronger and more uniform dipolar fields. Both wide and thin nano islands share the same length, resulting in poles (dark and bright lobes) positioned at equal distances. In contrast, the elongated leg in the asymmetric kagome ASI causes the poles to nearly overlap. Increasing the volume of kagome ASI enhances dipolar fields, thereby amplifying the nonlinearity of the system. Although tip charge contrast remains similar between narrow and asymmetric geometries, their vertex symmetry differs significantly. Narrow geometries maintain symmetric configurations, whereas asymmetric designs introduce frustration and imbalance, suppressing coherent magnetization dynamics. Coherent dynamics require symmetric, balanced dipolar fields. Breaking symmetry introduces frustration and imbalance, which destroys phase coherence of the collective precession. The key distinction lies in vertex dipolar field symmetry: as illustrated in table \ref{tab1b} synthetic MFM,  asymmetry disrupts the equilibrium among the three vertex legs, generating nonuniform internal fields and distorted vertex states. These asymmetric dipolar configurations impede coherent spin-wave propagation. Figure~\ref{fig2} presents the harmonic content generated by nonlinear 
scattering in the kagome lattice under a sinusoidal drive 
$\mathbf{h}(t) = h_{0}\sin(2\pi f t)\,\hat{x}$ with $f = 7$~GHz and 
$h_{0} = 2$--$10$~mT. Each run was integrated for 10~ns with the magnetization sampled every 4.46~ps. Panels~(a)--(f) of 
Fig.~\ref{fig2} show that the drive generates both integer harmonics 
and subharmonics of $f$.

As previously stated, the wide nano islands exhibit the highest harmonic and subharmonic amplitudes as a result of their dominant dipolar fields. In contrast, subharmonic amplitudes in thin and asymmetric nano islands are negligible.


Generation of the harmonics and subharmonics can also be used to excite the modes observed in the resonance spectra. For this purpose, we performed dedicated parametric excitation in which the applied field at $f=2f_\text{X}$ to probe the subharmonic response of the previously identified mode X, (cf.\ the mode positions in Fig.~\ref{fig1}(b)). The excitation of mode~X under $f=19.4,\;15.8,\;13.4\,$GHz for the 50, 80 and 110\,nm islands, respectively, is therefore indicative of parametric (parallel) pumping or the Suhl instability (see Fig.~\ref{figaa}). We observed that the applied field strength i.e., 
\( h_{0}=75,\;40,\;30\; \text{mT for 50, 80 and 110\,nm} \) required for the parametric excitation of mode X varies among the nano-islands due to variations in their demagnetization fields.   

Table \ref{tab2} identifies an additional mechanism by which geometric tip shape enhances nonlinearity in nanoisland arrays. In these islands the demagnetizing field is concentrated near the edges; therefore, rounding the tips into a dumbbell shape amplifies edge-localized demagnetizing fields and exchange fields , alters the local anisotropy landscape, and strengthens dipolar coupling between neighboring tips. Crucially, this increases tip–tip interaction strength without enlarging the overall island footprint or reducing inter-element spacing, offering a practical route to compact, strongly coupled arrays that remain fabrication-friendly. The curved tips also function as localized scattering centers, promoting mode mixing and enabling greater transfer of energy into higher-order modes. Figure~\ref{fig_e} quantifies the energetic consequences: the relaxed-state demagnetization energy changes modestly from $E_{\rm demag}=1.5245\times10^{-16}\,$J to $1.5422\times10^{-16}\,$J ($\approx 1.16\%$), whereas the exchange energy rises substantially from $E_{\rm exch}=1.1580\times10^{-17}\,$J to $1.4307\times10^{-17}\,$J ($\approx 23.6\%$). The near-constant total magnetostatic energy indicates a redistribution of stray fields, while the large increase in exchange energy signals enhanced local magnetization gradients near the rounded tips. Together, these effects increase nonlinear coupling and are consistent with the observed transition from predominantly harmonic emission in the sharp kagome islands to the appearance of subharmonics in the dumbbell geometry.

 In Fig. \ref{fig4} (a), the FFT amplitude spectra are shown as function of edge radius for dumbbell-shaped nano islands, highlighting the generation of harmonics and subharmonics. These spectra were computed for  $M_s=800$ kA/m $A=13$ pJ/m  and $\alpha = 0.01$, $\bold{h}\left(t\right)=h_{0} \sin \left(2 \pi f t\right)\hat{x}$, $f_{app}=7$ GHz and $h_{0}=2$ mT. As the edge radius increases, the spectrum changes from mostly harmonic response to clear subharmonic generation. That transition occurs because rounding the tips concentrates demagnetizing and exchange fields near the island ends, creating a localized nonlinear region. The important point is that this stronger nonlinear response appears without changing the island footprint or lattice spacing, so curvature acts as a footprint preserving control knob. Similarly, in Fig. \ref{fig4} (b) the effect of applied field strength is illustrated for  $f_{app}=9$ GHz. It is evident that harmonics and subharmonics are consistent above 2mT drive amplitude.

  \begin{figure}[!htph]
\centering
		\includegraphics[width=\linewidth]{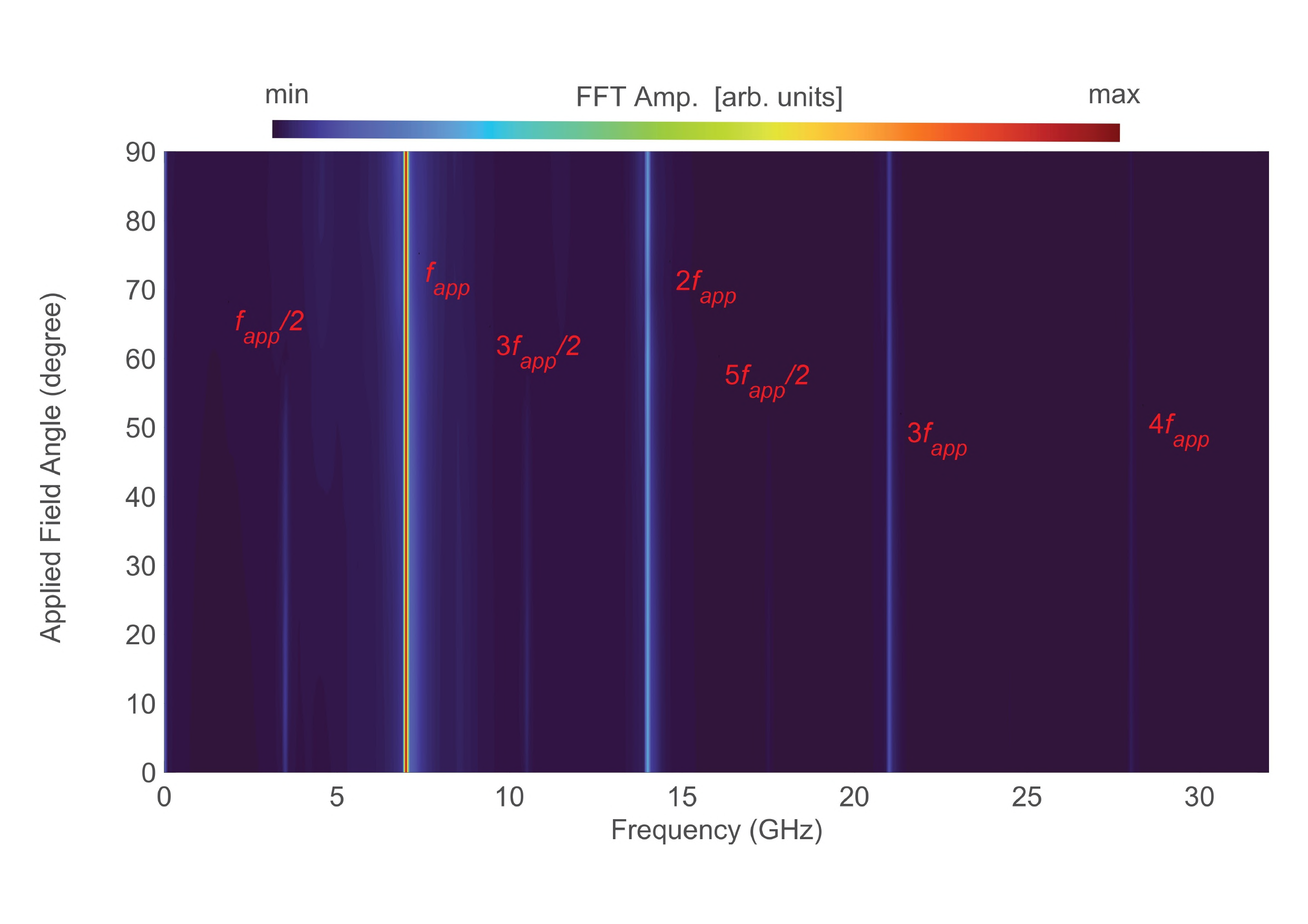}
	\caption{Angle–frequency map for dumbbell shaped kagome artificial spin ice under in plane excitation. Subharmonic branches appear over a finite angular window, showing that the nonlinear response depends not only on frequency and drive amplitude but also on field orientation relative to the island axis.}
	\label{fig6}
\end{figure} 
The nano islands possess a single geometric easy axis (a preferred in-plane magnetization direction set by shape anisotropy). In a kagome ASI the islands form three sublattices, so the array has three in-plane easy-axis directions separated by 120°. As a result, rotating the applied field changes its projection onto each island’s easy axis and therefore alters the effective switching field (coercivity) of different sublattices. This sensitivity is amplified when the island edge curvature (edge radius) is varied, since sharper or more rounded tips modify local demagnetizing and exchange fields. An example of this behavior appears in Fig. \ref{fig5} (a) showing hysteresis loops. These hysteresis loops are crucial as they maps the static energy landscape defining anisotropy, effective fields, and symmetry-breaking conditions that governs the nonlinear magnon interactions responsible for frequency multiplication. At 0$^\circ$ applied field, all islands are symmetric with respect to the field direction, so they switch almost simultaneously, giving a single large jump in the hysteresis loop with no steps. For other angles, the symmetry is broken, so different sublattices of islands experience different effective fields and reverse at different times. This sequential reversal produces steps in the hysteresis loop.

Figure~\ref{fig5} (b) shows the dependence of the second-harmonic amplitudes on the applied field direction for wide, thin, and dumbbell shaped nano islands of fixed length 260,nm. Here $0^\circ$ corresponds to the applied field along the island long axis ($\hat{x}$). The dumbbell geometry reaches its largest harmonic amplitudes at $0^\circ$, whereas thin islands increase monotonically from $0^\circ$ to $90^\circ$, and wide islands exhibit a non-monotonic maximum near $\approx40^\circ$. These trends reflect simple physical mechanisms. For thin (high-aspect) islands, rotating the field away from the long axis increases the transverse torque on the magnetization, producing larger-angle precession and stronger harmonic generation. Dumbbell islands concentrate demagnetizing fields at their curved ends; a field aligned with the long axis therefore most efficiently excites the localized edge dynamics associated with those ends, giving the largest response at $0^\circ$. Wide islands have weaker shape anisotropy but stronger inter-island dipolar coupling; at $\sim40^\circ$ the applied field geometry best balances the transverse torque and the dipolar bias among the three vertex legs, producing the strongest collective excitation. Hence, the applied-field orientation provides an effective knob to tune the efficiency of nonlinear scattering in these lattices. 

Similarly, Fig. \ref{fig6} shows the effect of the applied field direction on the harmonic generation of dumbbell shaped nano islands. These graph were computed for $\alpha = 0.01$, $\bold{h}\left(t\right)=h_{0} \sin \left(2 \pi f t\right)\hat{x}$, $f=7$ GHz, $R=112$ nm, and $h_{0}=2$ mT. Angle dependence of subharmonic generation is evident and is linked to the alignment of field with the field with the long axis so that it efficiently couples to the curved tip regions and their localized internal fields. The appearance of subharmonic branches over a finite angular window confirms that the nonlinear excitation is not only frequency dependent but also symmetry dependent.

\section{\label{sec:level5} Summary}
This work shows that the geometry of artificial spin ice can be used to control nonlinear magnon scattering through vertex level dipolar tuning. Using MuMax3 simulations, we studied kagome ASI structures with different nanoisland widths, leg asymmetry, and edge curvature. We found that wider nanoislands support stronger dipolar coupling and produce larger nonlinear harmonic signals than narrower ones. In contrast, asymmetric leg lengths distort the vertex configuration and increase magnetic frustration, which changes both the harmonic and subharmonic response. We also show that dumbbell shaped nanoislands, obtained by increasing edge curvature, enhance tip localized demagnetizing fields and promote higher order mode excitation. In addition, the orientation of the static applied field strongly affects coercivity and spectral content, making it another useful control parameter for nonlinear scattering. Together, these results demonstrate several geometric routes for tuning magnonic response in ASI, with edge curvature emerging as a particularly effective design knob.

\vspace{1cm}
This work was supported by the National Key R\&D
Program of China (Grants No. 2025YFA1411302), the
National Natural Science Foundation of China (Grant
No. 12374103 and 12434003, and Sichuan Science and Technology Program (No. 2025NSFJQ0045)

\bibliography{biblob.bib}

    

\end{document}